\RequirePackage{tikz} 
\documentclass[
twocolumn, 
twocolappendix,
numberedappendix,tighten,
]{aastex631mod}
\usepackage{amsmath}
\usepackage{amstext}
\usepackage[T1]{fontenc}
\usepackage{apjfonts} 
\usepackage{booktabs}
\usepackage{listings}
\usepackage[encapsulated]{CJK}
\usepackage{flowtex} 
\usepackage{xspace}
\usepackage[all]{hypcap}
\newcommand{\cntext}[1]{\begin{CJK}{UTF8}{gbsn}#1\end{CJK}}

\newcommand{\inprep}[1]{{\color{citecolor} #1 in prep.}}

\defcitealias{eimer23}{E23}

\newcommand{\marker}[3]{\tikz[baseline=-3pt]{
\definecolor{mycolor}{rgb}{#1,#2,#3}
\node[rectangle, scale=0.8, anchor=east,
rounded corners=1pt,
right=1pt of current bounding box.west,
minimum height=8pt,fill=mycolor](){}}\xspace}

\graphicspath{{./}{figures/}}
\definecolor{citecolor}{rgb}{0.08,0.30,0.85}
\hypersetup{citecolor=citecolor,filecolor=cyan}

\shorttitle{CLASS 40~GHz Maps}
\begin{document}
\title{CLASS  Data Pipeline and Maps for 40 GHz Observations through 2022}
\newcommand{\jhu}{The William H. Miller III Department of Physics and Astronomy, Johns Hopkins University, 3701 San Martin Drive, Baltimore, MD 21218, USA}
\newcommand{\ucsc}{Departamento de Ingenier\'{i}a El\'{e}ctrica, Universidad Cat\'{o}lica de la Sant\'{i}sima Concepci\'{o}n, Alonso de Ribera 2850, Concepci\'{o}n, Chile}
\newcommand{\villanova}{Department of Physics, Villanova University, 800 Lancaster Avenue, Villanova, PA 19085, USA}
\newcommand{\goddard}{NASA Goddard Space Flight Center, 8800 Greenbelt Road, Greenbelt, MD 20771, USA}
\newcommand{\uchicago}{Department of Astronomy and Astrophysics, University of Chicago, 5640 South Ellis Avenue, Chicago, IL 60637, USA}
\newcommand{\puci}{Instituto de Astrof\'isica, Facultad de F\'isica, Pontificia Universidad Cat\'olica de Chile, Avenida Vicu\~na Mackenna 4860, 7820436, Chile}
\newcommand{\pucc}{Centro de Astro-Ingenier\'ia, Facultad de F\'isica, Pontificia Universidad Cat\'olica de Chile, Avenida Vicu\~na Mackenna 4860, 7820436, Chile}
\newcommand{\argonne}{High Energy Physics Division, Argonne National Laboratory, 9700 S. Cass Avenue, Lemont, IL 60439, USA}
\newcommand{\upenn}{Department of Physics and Astronomy, University of Pennsylvania, 209 South 33rd Street, Philadelphia, PA 19104, USA}
\newcommand{\ucboulder}{Department of Astrophysical and Planetary Sciences, University of Colorado, 2000 Colorado Avenue, Boulder, CO 80309, USA}
\newcommand{\cfa}{Center for Astrophysics, Harvard \& Smithsonian, 60 Garden Street, Cambridge, MA 02138, USA}
\newcommand{\oslo}{Institute of Theoretical Astrophysics, University of Oslo, P.O. Box 1029 Blindern, N-0315 Oslo, Norway}
\newcommand{\MIT}{MIT Kavli Institute, Massachusetts Institute of Technology, 77 Massachusetts Avenue, Cambridge, MA 02139, USA}
\newcommand{\cepia}{CePIA, Astronomy Department, Universidad de Concepción, Casilla 160-C, Concepción, Chile}
\author[0000-0002-4820-1122]{Yunyang Li (\cntext{李云炀}\!\!)}
\affiliation{\jhu}
\correspondingauthor{Yunyang Li}
\email{yunyangl@jhu.edu}

\author[0000-0001-6976-180X]{Joseph R. Eimer}\affiliation{\jhu}
\author[0000-0003-2838-1880]{Keisuke Osumi}\affiliation{\jhu}
\author[0000-0002-8412-630X]{John~W. Appel}\affiliation{\jhu}
\author{Michael~K. Brewer}\affiliation{\jhu}
\author[0000-0001-7941-9602]{Aamir Ali}\affiliation{\jhu}
\author[0000-0001-8839-7206]{Charles L. Bennett}\affiliation{\jhu}
\author[0000-0003-2682-7498]{Sarah~Marie Bruno}\affiliation{\jhu}
\author[0000-0001-8468-9391]{Ricardo Bustos}\affiliation{\ucsc}
\author[0000-0003-0016-0533]{David T. Chuss}\affiliation{\villanova}
\author[0000-0002-7271-0525]{Joseph~Cleary}\affiliation{\jhu}
\author[0000-0002-0552-3754]{Jullianna Denes~Couto}\affiliation{\jhu}
\author[0000-0002-1708-5464]{Sumit Dahal}\affiliation{\goddard}\affiliation{\jhu}
\author[0000-0003-3853-8757]{Rahul Datta}\affiliation{\uchicago}\affiliation{\jhu}
\author[0000-0002-3592-5703]{Kevin~L. Denis}\affiliation{\goddard}
\author{Rolando D\"unner}\affiliation{\puci}\affiliation{\pucc}
\author[0000-0002-1052-0339]{Francisco Espinoza}\affiliation{\ucsc}
\author[0000-0002-4782-3851]{Thomas~Essinger-Hileman}\affiliation{\goddard}
\author[0000-0002-2061-0063]{Pedro Flux\'a Rojas}\affiliation{\puci}\affiliation{\pucc}
\author[0000-0003-1248-9563]{Kathleen Harrington}\affiliation{\argonne}\affiliation{\uchicago}
\author[0000-0001-7466-0317]{Jeffrey Iuliano}\affiliation{\upenn}\affiliation{\jhu}
\author{John Karakla}\affiliation{\jhu}
\author[0000-0003-4496-6520]{Tobias~A. Marriage}\affiliation{\jhu}
\author[0000-0002-2245-1027]{Nathan J.~Miller}\affiliation{\goddard}\affiliation{\jhu}
\author[0000-0001-7196-2520]{Sasha Novack}\affiliation{\ucboulder}\affiliation{\jhu}
\author[0000-0002-5247-2523]{Carolina N\'{u}\~{n}ez}\affiliation{\jhu}
\author[0000-0002-4436-4215]{Matthew~A.~Petroff}\affiliation{\cfa}
\author[0000-0001-5704-271X]{Rodrigo A. Reeves}
\affiliation{\cepia}
\author[0000-0003-4189-0700]{Karwan Rostem}\affiliation{\goddard}
\author[0000-0001-7458-6946]{Rui Shi (\cntext{时瑞}\!\!)}\affiliation{\jhu}
\author[0000-0003-3487-2811]{Deniz A. N. Valle}\affiliation{\jhu}
\author[0000-0002-5437-6121]{Duncan J. Watts}\affiliation{\oslo}
\author[0000-0003-3017-3474]{Janet L. Weiland}\affiliation{\jhu}
\author[0000-0002-7567-4451]{Edward J. Wollack}\affiliation{\goddard}
\author[0000-0001-5112-2567]{Zhilei Xu (\cntext{徐智磊}\!\!)}\affiliation{\MIT}\affiliation{\upenn}
\author[0000-0001-6924-9072]{Lingzhen Zeng}\affiliation{\cfa}
\collaboration{36}{CLASS Collaboration}
\received{2023 May 1}
\revised{2023 July 26}
\accepted{2023 August 20}
\submitjournal{\apj}
\begin{abstract}
The Cosmology Large Angular Scale Surveyor (CLASS) is a telescope array that observes the cosmic microwave background over 75\% of the sky from the Atacama Desert, Chile, at frequency bands centered near 40, 90, 150, and 220~GHz. 
This paper describes the CLASS  data pipeline and maps for 40~GHz observations conducted from 2016 August to 2022 May.
We demonstrate how well the CLASS survey strategy, with rapid ($\sim10\,\mathrm{Hz}$) front-end modulation, recovers the large-scale Galactic polarization signal from the ground: the mapping transfer function recovers $\sim67\%\,(85\%)$ of $EE$ and $BB$ ($VV$) power at $\ell=20$ and $\sim35\%\,(47\%)$ at $\ell=10$.  We present linear and circular polarization maps over $75\%$ of the sky. Simulations based on the data imply the maps have a white noise level of $110\,\mathrm{\mu K\, arcmin}$ and correlated noise component rising at low-$\ell$ as $\ell^{-2.4}$. The transfer-function-corrected low-$\ell$ component is comparable to the white noise at the angular knee frequencies of  $\ell\approx18$ (linear polarization) and $\ell\approx12$ (circular polarization). Finally, we present simulations of the level at which expected sources of systematic error bias the measurements, finding subpercent bias for the $\Lambda$ cold dark matter $EE$  power spectra. Bias from  $E$-to-$B$ leakage due to the data reduction pipeline and polarization angle uncertainty approaches the expected level for an $r=0.01$ $BB$ power spectrum. Improvements to the instrument calibration and the data pipeline will decrease this bias. 
\end{abstract}

\keywords{ 
    \href{http://astrothesaurus.org/uat/435}{Early Universe (435)}; 
    \href{http://astrothesaurus.org/uat/322}{Cosmic microwave background radiation (322)};  
    \href{http://astrothesaurus.org/uat/1146}{Observational Cosmology (1146)}; 
    \href{http://astrothesaurus.org/uat/1277}{Polarimeters (1277)}; 
    \href{http://astrothesaurus.org/uat/1858}{Astronomy Data Analysis (1858)}}
\section{Introduction}
The quest for a precise understanding of cosmology has propelled the development of cosmic microwave background (CMB) observations. Satellite missions like COBE, \citep{boggess1992cobe}, WMAP \citep{bennett03,bennett13,hinshaw13}, and Planck 
\citep{tauber2004planck,planck18I} have made all-sky measurements of the CMB anisotropy in 
temperature and polarization, which are cornerstones supporting the standard model 
of cosmology. Related ground-based and balloon-borne efforts provide first-look results, cross-checks, and extended capabilities (e.g., higher resolution and/or higher sensitivity) to the satellites. Suborbital efforts also develop new technologies (e.g., cryogenic detectors and high-throughput optics) to space-readiness levels and train the next generation of experimental cosmologists. 
These experiments typically observe patches of the sky ranging from degree scale
to $\sim40\%$ of the sky \citep[e.g.,][]{Aiola20}.
Over the past decade, suborbital observations have offered complementary views of the CMB and 
tightened the constraints on cosmological parameters through
improved measurement of the damping tail of the temperature power spectrum \citep{choi20,dutcher21},
the linear polarization spectra at and below $\sim 5^\circ$ angular scales  \citep{pb20bmode,BK-XV20,spider21}, 
the gravitational lensing potential \citep{ade14,BK-VIII16,bianchini20,madhavacheril23},
deep surveys of galaxy clusters \citep{bleem20, hilton21},
and characterization of the Galactic foregrounds \citep{quiet15, cbass22, quijote23}.

However, ground-based CMB polarimetry has been largely unexplored on the largest angular scales \citep[$\ell\lesssim30$,][]{quiet15, BK16, abs18} due to fluctuations in atmospheric emission and other sources of systematic error arising from the interaction of the telescope with its environment.
This has become an impediment to the percent-level characterization of the reionization history of the universe \citep{zaldarriaga97,pagano19,watts20} and to the search for tensor perturbations on the largest angular scales \citep{guth81,kamionkowski97,seljak97,Tristram21-PR4-r}. 
While the search for tensor perturbations has progressed at $\ell\gtrsim30$ led by \cite{BK21}, the largest-scale $B$ modes would provide the distinctive ``reionization peak'' feature and would be most
unambiguously separable from the late-time lensing effect \citep{zaldarriaga98}.
The largest angular scales also provide access to beyond-the-standard-model physics \citep[e.g.,][]{muir18,hogan19,hogan23,shi23} and the physics of the interstellar medium \citep[e.g.,][]{caldwell17}. 
It is the goal of the Cosmology Large Angular Scale Surveyor (CLASS) project to develop the technology and techniques needed to explore the large-scale CMB polarization from the ground.

CLASS is a telescope array located in the Atacama Desert in Chile \citep{essinger-hileman14spie} that observes at frequencies around 40, 90, 150, and 220~GHz and surveys
75\% of the sky every day.
Access to the largest angular scales is enabled through rapid front-end modulation with a variable-delay polarization modulator \citep[VPM;][]{chus12vpm,harrington18spie,harrington21}, which also suppresses instrumental polarization.
Compared to other modulation technologies, such as 
the half-wave plate \citep[][]{abs14,pb17hwp}, the employment of a VPM enables 
CLASS's unique sensitivity to circular polarization \citep{petroff20,padilla20}. 
CLASS also employs boresight rotations, an optical design that prioritizes signal fidelity over maximizing throughput \citep{eimer12spie,xu20,datta23}, and a fully enclosed comoving ground shield, to map the largest angular scales. CLASS measurements complement large-scale data from future satellite \citep{litebird22} and balloon-borne \citep{piper14,taurus20apra,lspe20} telescopes as well as other ground-based strategies \citep{groundbird20,lspe20, quijote23}. Major upcoming international-scale ground-based surveys target scales $\ell\gtrsim 30$ \citep{simons19whitepaper,stagefour22bmodeforecast}.

In this paper, we describe the data reduction pipeline and polarization maps of the CLASS $40\,\mathrm{GHz}$ data
taken through 2022. 
Angular power spectra and other map-based results are presented in a companion paper \citep[hereafter E23]{eimer23}. The paper is organized as follows. 
Section~\ref{sec:overview} overviews the design of the 40~GHz telescope and the survey. 
Section~\ref{sec:data-processing} explains the main data processing steps. 
Mapmaking and the survey maps are presented in Section~\ref{sec:map-making}.
The impact of several systematic issues is reviewed in Section~\ref{sec:systematics}.
The Appendix provides a description of the pointing model.

\section{Overview}\label{sec:overview}
\begin{figure*}
\centering
\includegraphics[width=\linewidth]{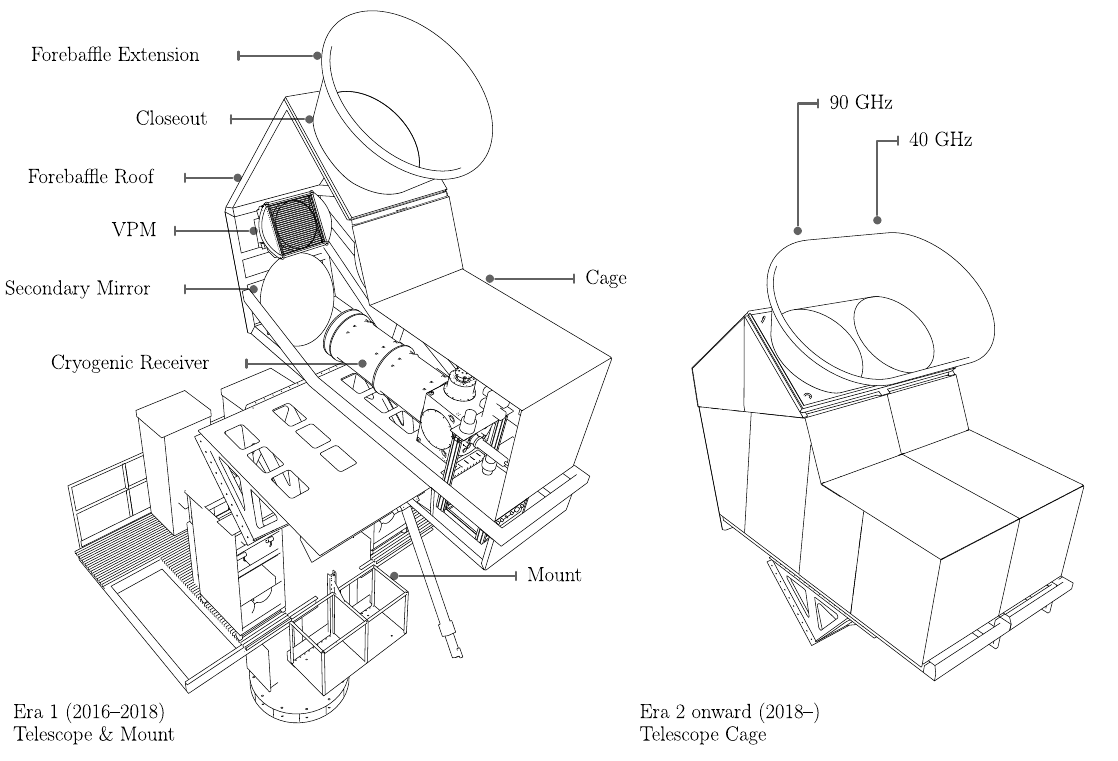}
\caption{Schematic of the CLASS 40~GHz Telescope. The telescope sits atop the three-axis telescope mount in an enclosed cage structure. Lower-side cage panels are not shown at left to reveal the telescope. Light enters the enclosure by the forebaffle extension, first encountering the VPM, then the primary (obscured) and secondary mirrors, and finally the receiver. The enclosed multireflection design limits spurious signals from stray light. In 2018, the enclosure and forebaffle extension were extended to accommodate the 90~GHz telescope on the same mount.
}\label{fig:telescope-sketch}
\end{figure*}

\subsection{The CLASS 40 GHz Telescope}
\label{ssec:telescope}

The CLASS telescope array is sited on Cerro Toco at 5140 m in the Atacama Desert of northern Chile (latitude $-22.96^\circ$, longitude $-67.79^\circ$),
a location long recognized for its microwave-transparent atmosphere due to the combination of high elevation and low precipitable water vapor \citep[PWV;][]{cortes20}. During the observations presented in this paper, PWV quartiles were $[0.63, 1.10, 1.98]$~mm.\footnote{CLASS PWV data are obtained from the APEX weather station: \url{https://www.apex-telescope.org/ns/weather-data}.} Given its proximity to the equator, the site also provides access to most of the sky, which is essential for large-angular-scale measurement.

Figure~\ref{fig:telescope-sketch} shows schematics for the 40~GHz telescope, both in its single-telescope configuration (2016--2018; left panel) and when it was paired with the 90 GHz telescope (2018--present; right panel). The telescope is shown on its mount structure, which includes three axes of rotation: azimuth, elevation, and boresight. 
The boresight axis is aligned with the center of the telescope's field of view and has a full range of motion from $-45^\circ$ to $45^\circ$ with respect to a nominal central position. With the boresight axis so defined, the azimuth and elevation coordinates give the direction of the boresight axis.
Since only one linear polarization state is modulated by the VPM at a time, execution of boresight rotation is needed for a full sampling of the linear polarization signal. 
The elevation axis allows for $26^\circ-86^\circ$ rotation, but the VPM drive system restricts polarized observations to $40^\circ-55^\circ$. The azimuth has a $720^\circ$ range centered on the geographic south. Atop the mount, the cryogenic receiver is secured to a baseplate. Supporting instrumentation, including the helium compressor, gas handling system, and drive electronics sit on a platform that moves with the telescope in azimuth, simplifying the routing of cables and hoses for the large azimuth scans. An aluminum cage structure rises above the receiver and supports the telescope mirrors and the VPM. Aluminum honeycomb panels are bolted to the cage to enclose the telescope, blocking radiation paths from the ground. For the majority of observations analyzed here, the aluminum panels were coated on the inside by microwave absorbers (Eccosorb HR-10). Light enters the  cage enclosure through an extension cone, which is rolled at the top to decrease diffraction from sources away from the telescope boresight. 
We refer to the whole upper part of the cage enclosure above the VPM as a \emph{forebaffle} as it serves to reject incoming stray light through reflection or absorption. This consists of two parts: the \emph{forebaffle roof} and the \emph{forebaffle extension} as labeled in Figure~\ref{fig:telescope-sketch}.

To accommodate the 90~GHz telescope on the same mount in 2018, the cage structure and forebaffle were expanded as shown in the right side of Figure \ref{fig:telescope-sketch}. 
Other notable changes were implemented at this time such that we designate the time before  2018-02-22 as {\it Era~1} and after 2018-06-22 as {\it Era~2}. See Section \ref{ssec:changes} for further discussion. Era~1 has a total of 540~days (1.48~yr). Era~2, which was interrupted by the pandemic, has 1038~days (2.84 yr).

After light passes the forebaffle, it first encounters the VPM. 
Positioning the VPM as the first optical element modulates the incoming polarization such that it can later be separated from instrumental polarization, which is not modulated \cite[e.g.,][]{miller16}.  
After the polarization state is encoded by the modulator, the signal is guided by two ambient-temperature mirrors into a cryogenic receiver. The receiver uses a horizontally mounted Bluefors\footnote{\url{bluefors.com}} pulse-tube-backed helium dilution refrigerator that continuously cools the receiver optics and focal plane \citep{iuliano2018spie}. 
It employs a combined strategy of multilayer foam, reflective metal mesh, and plastic absorptive filters to block infrared radiation. Plastic lenses produce a telecentric image of the sky on a feedhorn-coupled detector array at the focal plane. The telescope's average beam FWHM is $1.54^\circ$ \citepalias{eimer23}, and its field of view spans $20^\circ$ in the azimuth direction and $15^\circ$  in elevation for zero boresight rotation. The absolute polarization angle calibration will be discussed in Section \ref{ssec:sys-pol-angle}.

In the focal plane, smooth-walled feedhorns \citep{zeng10spie} couple incoming radiation to 36 detector chips. On-chip ortho-mode transducers cleanly separate the $\pm45^\circ$ linear polarization states, the power levels of which are detected by separate transition-edge-sensor (TES) bolometers \citep{chuss12tes}. Therefore, the telescope has 72 feedhorn-coupled TES bolometers (an orthogonally polarized pair for each horn/chip).
Bandpass filters on the detector chip define a frequency band centered on 38~GHz with a 12.3~GHz bandwidth \citep{dahal22}. A cryogenic superconducting quantum interference device (SQUID) based time-division multiplexing architecture and ambient-temperature Multi-Channel Electronics (MCE) read out the detectors \citep{reintsema03,battistelli08}.
Details regarding the array construction and performance can be found in \cite{rostem12spie}, \citet{appel14spie,appel19}, and \cite{dahal22}. 
The dilution refrigerator cools the entire focal plane to $\sim 40\,\mathrm{mK}$, which serves as the base temperature for the bolometers that are voltage-biased at their 150~mK transition temperature. 
At these low temperatures, the primary source of bolometric noise is from the stochasticity of the incoming radiation load \citep[$1.2\,\mathrm{pW}$ median;][]{appel19,dahal22}. For Era~1, the nominal detector array sensitivity\footnote{The referenced numbers correspond to the noise-equivalent temperature of the raw data in the white noise regime. 
With the VPM modulation, the polarization sensitivities are degraded by a factor corresponding to the (inverse of) \textit{modulation efficiency}, which is around 0.72/0.48 for linear/circular polarizations, respectively (see Eq. \ref{eq:data-model}). } was $32\,\mathrm{\mu K\sqrt{s}}$. 
Changes to the instrument described in \cite{dahal22} increased the number of working detectors, improved optical efficiency, and reduced optical loading and resulted in a decrease to $30.5\,\mathrm{\mu K \sqrt{s}}$ in Era~2.

\subsection{Observations}\label{ssec:survery-overview}
\begin{figure*}
    \includegraphics[width=\linewidth, ]{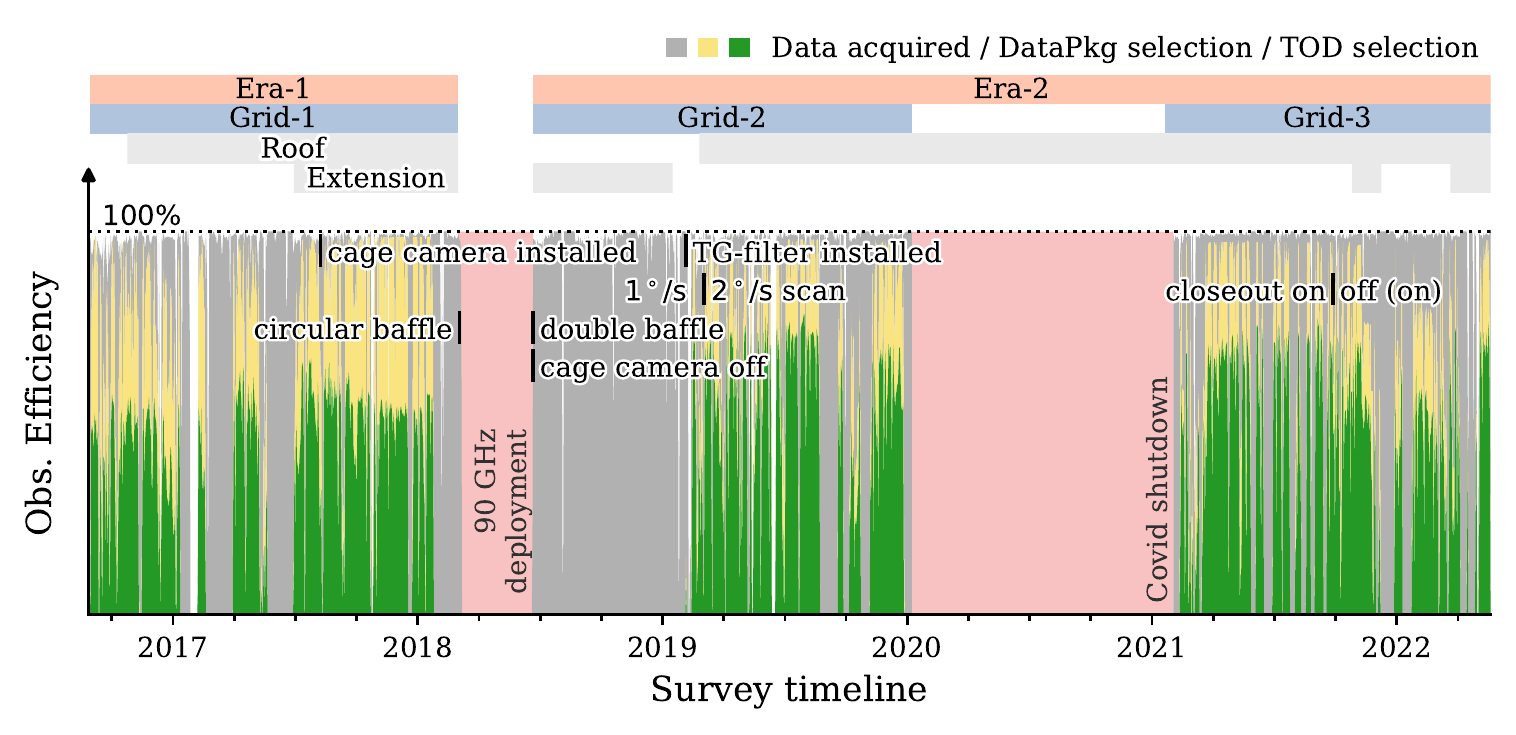}
    \caption{
    CLASS daily observation efficiencies from 2016-08-31 to 2022-05-19. 
    The gray region is the total detector time recorded when the VPM is working; the yellow region shows the amount of data initially selected at the DataPkg level (Section \ref{ssec:dpkg-selection}), and the green region shows the fraction 
    after the TOD-level selection that is used for mapping (Section \ref{ssec:tod-selection}).
    The time spans for Era~1 and Era~2 are shown in orange at the top; three periods with different VPM grids are shown in blue.
    The gray hatched regions indicate times when different parts of the forebaffle are blackened.
    Critical changes to the instrument configuration are marked by the vertical bars, with the text pointing at the direction where the annotation applies.
    The closeout was first taken off on 2021-09-27 but was occasionally reinstalled to guard against bad weather.
    }
    \label{fig:class-overview}
\end{figure*}

This paper covers 40~GHz observations from their beginning on 2016-08-31 through 2022-05-19 (2089 days; 5.72 yr). Nominally, the CMB survey observations were conducted 24 hr per day with no restriction on the time of year. In practice, scheduled calibrations, maintenance, and instrument upgrades as well as unscheduled weather and other events interrupted the survey.  Figure \ref{fig:class-overview} shows the survey observation efficiency over time with 100\% indicating nominal 24-hr-per-day operation. Two major stoppages totaling 502 days are shown in pink fill. These were due to the addition of the 90 GHz telescope on the same mount in 2018 and to a VPM grid failure in 2020 that was not repairable until COVID travel restrictions were eased a year later. Because these did not have to do with the regular operation of the 40~GHz telescope, we excluded the associated time when estimating the total possible data volume and observing efficiency. With this exclusion, the total time under consideration is 1587~days (4.32~yr), which is divided between Era 1 (1.48 yr) and Era 2 (2.84~yr) as described in Section~\ref{ssec:telescope}.  In addition to the two major stoppages shown in pink, a number of other longer periods (not demarcated in Figure \ref{fig:class-overview}) were excluded. This could be for sustained bad weather, such as the two roughly month-long periods in the austral fall and winter of 2017, or for systematic instrument malfunction such as after the 90~GHz deployment through early 2019 when radio-frequency interference (RFI) compromised the 40~GHz data. After these exclusions, 64.6\% of the data remained (Table \ref{tab:data-selection}, ``Timeline''). Data quality cuts reduced the usable data volume further and will be discussed in Section~\ref{sec:data-processing}.

\begin{figure}
    \centering
    \includegraphics[width=\columnwidth]{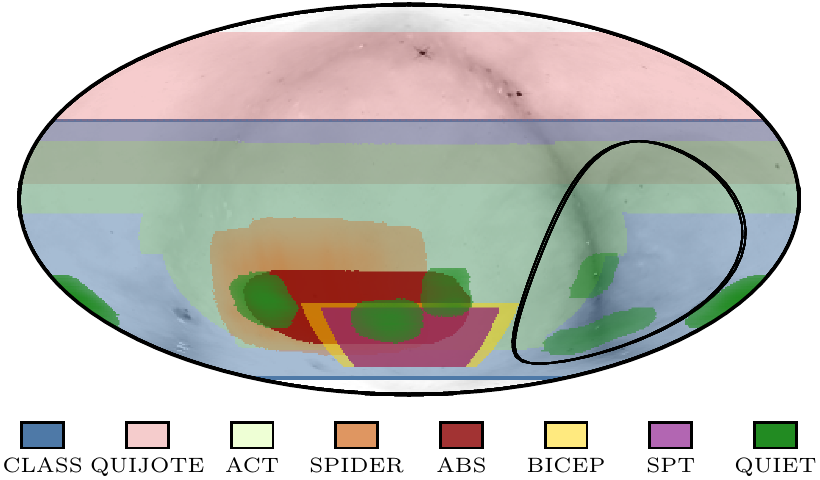}
    \caption{CLASS survey footprint in equatorial coordinates. CLASS observes the sky with large constant-elevation circular scans, illustrated by the black curve that traces the path of the telescope boresight through one $720^\circ$ sweep. The thickness of the curve in R.A. (separable into two lines upon zooming in) indicates the degree of sky rotation during a sweep. Over a 24 hr period, the sweeps cover the annular survey region shown between the dark blue lines, extending from decl. $+30^\circ$ in the north to $-76^\circ$ in the south. 
    The background image shows the intensity of the Planck component-separated synchrotron map \citep{planck18IV}. Survey footprints from a sampling of other projects are superposed \citep{quijote23, Aiola20,spider21, abs18, BK-XV20, dutcher21, quiet15}.}
    \label{fig:footprints}
\end{figure}
 
During CMB survey observations, the telescope scanned in azimuth with the elevation and boresight angles fixed. With few exceptions, the elevation was held at $45^\circ$. Each day the telescope boresight rotation was set at an angle between $-45^\circ$ and $45^\circ$ in $15^\circ$ increments; the seven boresight angles were visited once a week during observations. The azimuth scan covered the full $ 720^\circ$ range, centered on the geographic south ($180^\circ$). Therefore, the telescope traveled from $-180^\circ$ to $540^\circ$ (a forward {\it sweep}) and then back from $540^\circ$ to  $-180^\circ$ (a backward {\it sweep}). This simple back-and-forth circle scan was repeated throughout the observations. Because the telescope is located at latitude $-23^\circ$, during each sweep the boresight traced out a large circle on the celestial sphere centered on decl. $-23^\circ$ with a $45^\circ$ radius. Accounting for the footprint of the field of view about the boresight pointing, the CLASS data extend from decl. $30^\circ$ in the north to $-76^\circ$ in the south. As the Earth rotates, the circle scans swept through the full 24 hr of R.A. to cover 75\% of the sky. Figure \ref{fig:footprints} shows the CLASS survey region in celestial coordinates overlaid on the Planck synchrotron temperature map \citep{planck18IV}. The footprints of other CMB surveys are also shown. 

One exception to the simple circle scan described above occurred during daytime observations when the Sun passes over the scan path. The telescope maintained a separation greater than $20^\circ$ between the boresight and the Sun position. Therefore, the scans were truncated to less than $360^\circ$ azimuth as the Sun rose or set through the scan path.

\subsection{Changes to the Instrument and Observations}
\label{ssec:changes}

Several adjustments in instrument configuration or observing strategy have taken place since the beginning of the survey, many of which were due to instrument updates, replacement of failed components, and remedies for systematics guided by analysis. In Section \ref{ssec:telescope}, we already noted the enlargement of the cage and forebaffle extension to accommodate the 90~GHz telescope---alterations that demarcate parts of the survey are labeled Era 1 (before 2018-02-22) and Era 2 (after 2018-06-22). Important changes are noted in Figure \ref{fig:class-overview} and explained below in approximate time order. 

\paragraph{Cage Camera Interference} On 2017-08-09, webcams were installed inside the cage enclosure to monitor the telescope. 
These cameras were later found to introduce RFI around the CLASS modulation frequencies and were turned off for CMB observation from 2018-06-22 onwards. The effects in demodulated data are at frequencies below the scanning frequency and have little impact on the result; see Section~\ref{ssec:camera-signal}.

\paragraph{Forebaffle Roof Blackening and Geometry Change} 
To improve the rejection of stray light, microwave absorbers (Eccosorb HR-10) were attached to the inside top and walls of the forebaffle roof first on 2016-10-25. When the forebaffle roof was replaced in Era 2, it was not only extended to accommodate the 90~GHz telescope; its geometry (e.g., angle of the roof) was also changed to improve the rejection of stray light. Eccosorb HR-10 was also applied to the inside of the new forebaffle roof on 2019-02-25. (The data taken while the new Era-2 forebaffle roof was unblackened were rejected due to RFI; see below, unrelated to the blackening.) The time during which the forebaffle roof was blackened is hatched in Figure~\ref{fig:class-overview}. 

\paragraph{Forebaffle Extension Blackening} 
To prevent fractionally ($\sim10^{-3}$) polarizing reflections seen in detectors toward the edge of the field of view during moon observations \citep[Figure 19 of][]{xu20}, the forebaffle extension interior was first blackened with a microwave absorber (Eccosorb HR-10) on 2017-07-20. The blackening was retained through the end of Era 1 and attached to the new baffle at the beginning of Era~2. After suffering damage, the blackening was removed on 2019-01-16 and not replaced until late in Era~2. 
The time of the survey when the forebaffle extension was blackened is hatched in Figure~\ref{fig:class-overview}.
See also Section  \ref{ssec:az-signal}.

\paragraph{Variable Speed Azimuth Scan Test}  From 2017-04-01 to 2017-05-18, a variable scan speed strategy (\emph{dec scan}) was explored to even the integration time across the decl. range. This was abandoned for constant velocity scans (\emph{az scan}) when detector-warming vibrations were induced at certain velocities explored by the dec scans.  

\paragraph{Focal Plane Fix} The initial deployment of the 40~GHz receiver had 64 of 72 optical detectors working, 56 of which were in pairs. The inoperable channels were due to multiplexer failures, likely from static discharge on address lines.
The receiver updates for Era 2 recovered all optical detectors.

\paragraph{VPM Grid 2} The Era-1 VPM wire grid had brown oxidation on its wire grid and imperfections toward its lower edge. Suspecting these may be responsible for larger-than-expected grid emissions, a new grid was installed for Era 2. However, no significant change in performance was observed. 

\paragraph{Infrared Filter Changes, RFI, and Thin Grille Filter} Infrared filtering changes made between Era 1 and Era 2 increased the telescope's optical efficiency such that the noise-equivalent temperature (for both white and correlated components) dropped by 20\%. However, additional RFI, either due to a new Era 2 component (e.g., new VPM drive electronics) or increased susceptibility of the receiver, required the introduction of a warm thin grille filter (TG-filter) at the receiver window on 2019-01-12. This canceled the efficiency gains from the increased in-band transmission of the new infrared filters \citep[][\inprep{Cleary et al.}]{dahal22}.

\paragraph{Azimuth Scan Speed Increase} The
az-scan speed was increased from $1$ to $2\,\mathrm{deg}\,\mathrm{s}^{-1}$ on 2019-03-04. 
This was found to decrease the low-frequency noise in the demodulated data at 40~GHz (\inprep{Cleary et al.}). 
This indicates that the correlated noise component may be due to the residual ground signals after filtering (Section \ref{ssec:map-filter}) and/or the temporal correlations in the turbulent atmosphere emission \citep{morris2021} leaking into polarization; in the latter case, the faster signal modulation would permit better separation of the correlated noise from the sky signal.

\paragraph{VPM Grid 3} The second VPM grid failed on 2020-01-08, likely due to heating of the grid-securing epoxy during exposure to direct sunlight. Delayed by pandemic travel  restrictions, a third VPM grid was installed, and observations resumed on 2021-02-11.

\paragraph{Closeout Removal} A thin ($17.8\,\mathrm{\mu m}$) plastic environmental seal, the ``closeout'' in Figure \ref{fig:telescope-sketch}, was used where the light enters at the base of the forebaffle extension (diameter ${\sim}1.3\,\mathrm{m}$). 
We found that the closeout produced polarization systematics when deformed by the wind (Section \ref{ssec:wind-signal}). 
Since 2021-09-27, the closeout has been removed for most CMB observations and has only been temporarily reinstalled during bad weather. 

\section{Data Processing and Demodulation}\label{sec:data-processing}

The CLASS data reduction pipeline is designed to ingest the raw data and output well-characterized maps and spectra.
The use of the VPM for signal modulation naturally divides the pipeline into two parts.
First, raw detector time-ordered data (TOD) are vetted, calibrated,  demodulated, and downsampled into linear and circular polarization TOD. 
The subsequent mapmaking pipeline identifies and removes polarization systematics and solves for sky maps given the demodulated TOD.
Figure~\ref{fig:pipeline-flowchart} provides an overview of this pipeline.
In this section, we describe the procedures up to and including the demodulation and defer the discussion of the rest of the pipeline to Section~\ref{sec:map-making}.

\begin{nolinenumbers}
\begin{figure*}
\centering
\input{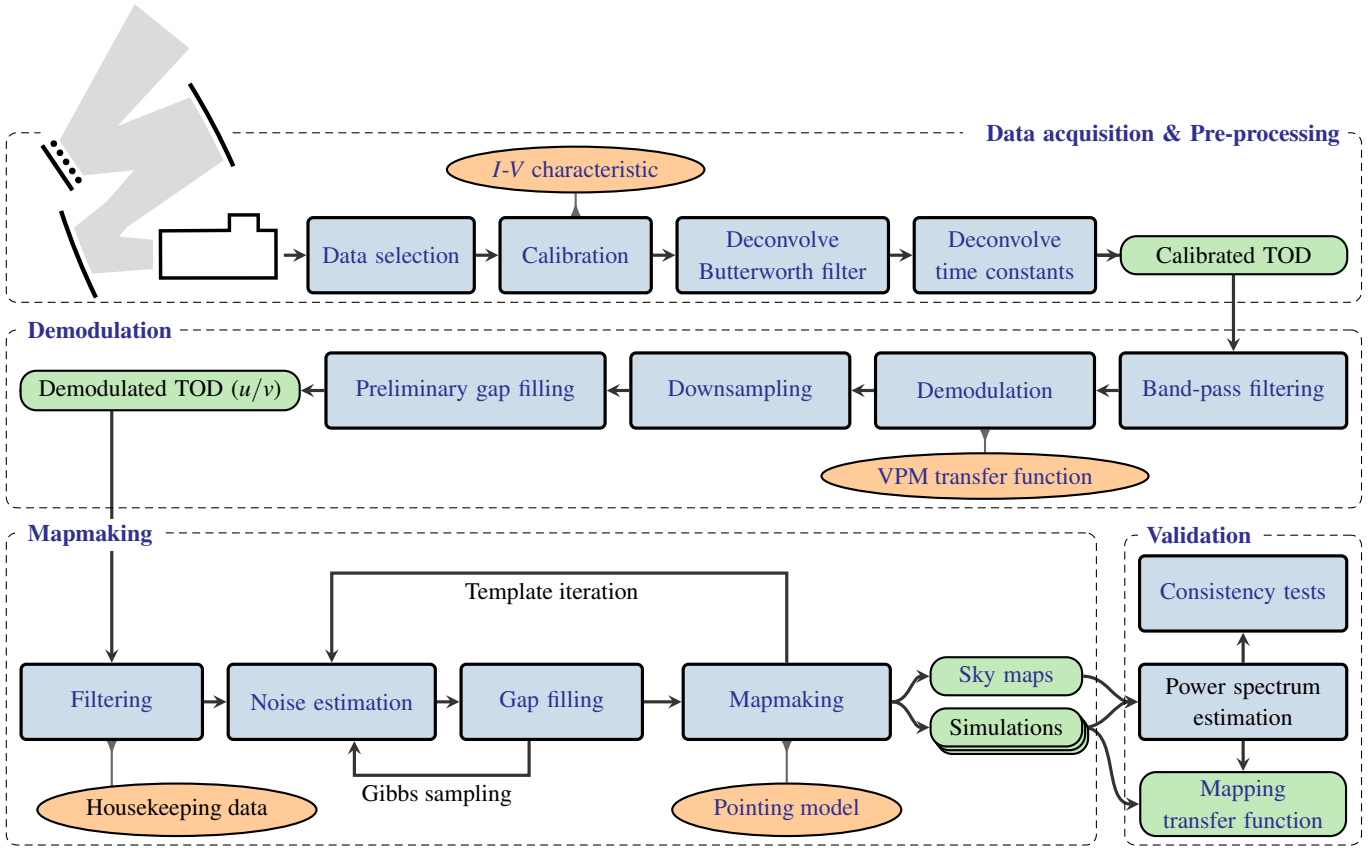}
\caption{Schematic of the CLASS data reduction pipeline. 
Key processing steps, data products, and ancillary input data are labeled as blue boxes, green pills, and orange ellipses, respectively. 
Whenever applicable, the entries in the chart are hyperlinked to the sections in this paper or external references where details are supplied.}
\label{fig:pipeline-flowchart}
\end{figure*}
\end{nolinenumbers}

\subsection{Data Acquisition and Selection}\label{ssec:dpkg-selection}
The detector data and the encoder data from the telescope mount and the VPM are measured synchronously with a rate of $201$ samples per second and grouped into clock-aligned 10 minute packages. 
These data, in combination with various asynchronously collected housekeeping data, are collated and saved on disk as \emph{data packages} (DataPkg), which are the smallest units for data storage and transmission. \citep[See][for details of the data acquisition pipeline.]{petroff20spie}

A total of $224,781$ DataPkgs ($5\,\mathrm{TB}$, in compressed volume) were collected for  40~GHz during the 4.32~yr of observations considered here. These are represented by the gray area in the timeline of Figure \ref{fig:class-overview}. As discussed in Section \ref{ssec:survery-overview}, several periods of sustained poor weather and instrument malfunction were discarded, leaving 64.6\% of the total data (Table \ref{tab:data-selection}, ``Timeline''; unmarked in Figure \ref{fig:class-overview}).
After this initial exclusion, an additional downselection was performed based on the metadata of each DataPkg.
Data acquired during unfavorable conditions (instrument maintenance, short-term poor weather, etc.) were dropped.
Furthermore, only the DataPkgs collected when the mount elevation was above $40^\circ$  were kept for processing.
As a final step,  DataPkgs were discarded if the cloud cover as monitored by the site webcams \citep{Li23cloud} was consistently too high to avoid highly variable data triggered by enhanced optical loading or polarization from clouds \citep{Takakura2019}.
These data cuts left $102,003$ ($45.4\%$) DataPkgs (Table \ref{tab:data-selection}, second row); their distribution is shown in yellow in Figure \ref{fig:class-overview}. 

The selected contiguous DataPkgs were concatenated to form \emph{spans} for data processing. 
A span typically has $\sim130$ DataPkgs (22 hr) of data, interrupted every day at around noon when the boresight angle was incremented by $15^\circ$  and detector gain calibrations were performed. However, spans may also be shorter if observations were interrupted (e.g., for planet observations).

\subsection{TOD Selection}\label{ssec:tod-selection}
While the previous section described data cuts that removed entire DataPkgs, additional analysis was performed on the resulting spans to flag data samples that are of low quality or susceptible to systematic issues.
The second half of Table \ref{tab:data-selection} enumerates these flags along with the fraction of data retained at each step. After the TOD selections, there remained 28\% of the total data volume, corresponding to 86.77 detector$\cdot$years of data for mapmaking. These data are represented in green in the timeline of Figure \ref{fig:class-overview}. A description of each TOD selection step is given below.

\paragraph{Survey Interruption} Over the course of a span, any incidental interruptions to the nominal CMB survey, e.g., for targeted source observations, were flagged. Similarly, interruptions to key instrument system performance were flagged. Monitored systems included the VPM, the full cryostat health, the telescope mount, and the detector readout system. Environmental factors were also monitored; in particular, data taken when the PWV exceeded 5~mm were flagged. This threshold was selected as a pragmatic limit by which strong atmospheric effects were avoided with only a modest impact on overall observing efficiency.

\begin{deluxetable}{llr}
\tablecaption{Data Selection and Processing Flags  \label{tab:data-selection}}
\tablehead{\colhead{Category} & \colhead{Item} & {Retained Data}}
\startdata
\specialrule{.1em}{0em}{0em}
DataPkg selection
& Timeline            &64.6\%\\
& Initial selection   &45.4\%\\
\cline{2-3} 
Subtotal &               &102,003 DataPkg\\
\specialrule{.08em}{0em}{0em}
TOD selection \tablenotemark{*}
& Survey interruption             & 95.5\%\\
& Transient detector cuts         & 88.1\%\\
& Source avoidance                & 70.3\%\\
& VSS amplitude cuts              & 65.1\%\\
& Conditioning cuts               & 62.1\%\\
\cline{1-3} 
Total  &                &86.77 detector$\cdot$years\\
\specialrule{.1em}{0em}{0em}
\enddata
\tablenotetext{*}{The percentages are the remaining fraction from the data package selection.}
\tablecomments{The quoted retained fraction at each step depends on the order in which selections are applied.}
\end{deluxetable}

\paragraph{Transient Detector Cuts} The detector TOD were analyzed to flag periods of anomalous behavior. Data from detectors with constant output were flagged. 
Data for which the SQUID flux-locked closed-loop detector readout became unstable were flagged. 
To recover these unlocked detectors and reestablish the SQUID tuning state optimized at the start of each span, the first stage SQUID feedback for all detectors was occasionally relocked. 
The data at the time of relock were flagged. 
Additionally, the SQUID readout can experience sudden jumps that manifest as discontinuities in the detector data; such jumps were flagged. 
During any window of 100 samples ($0.5\,\mathrm{s}$), if the total array of detectors experienced more than 10 jumps, all detectors were flagged for that window. Finally, the 100 samples ($0.5\,\mathrm{s}$) surrounding the azimuth drive direction reversals were flagged. Further information regarding the operation of the SQUID multiplexing amplifiers can be found in \cite{reintsema03} and \cite{nist_tdm_mux13b}.

\begin{figure}
\includegraphics[width=\linewidth]{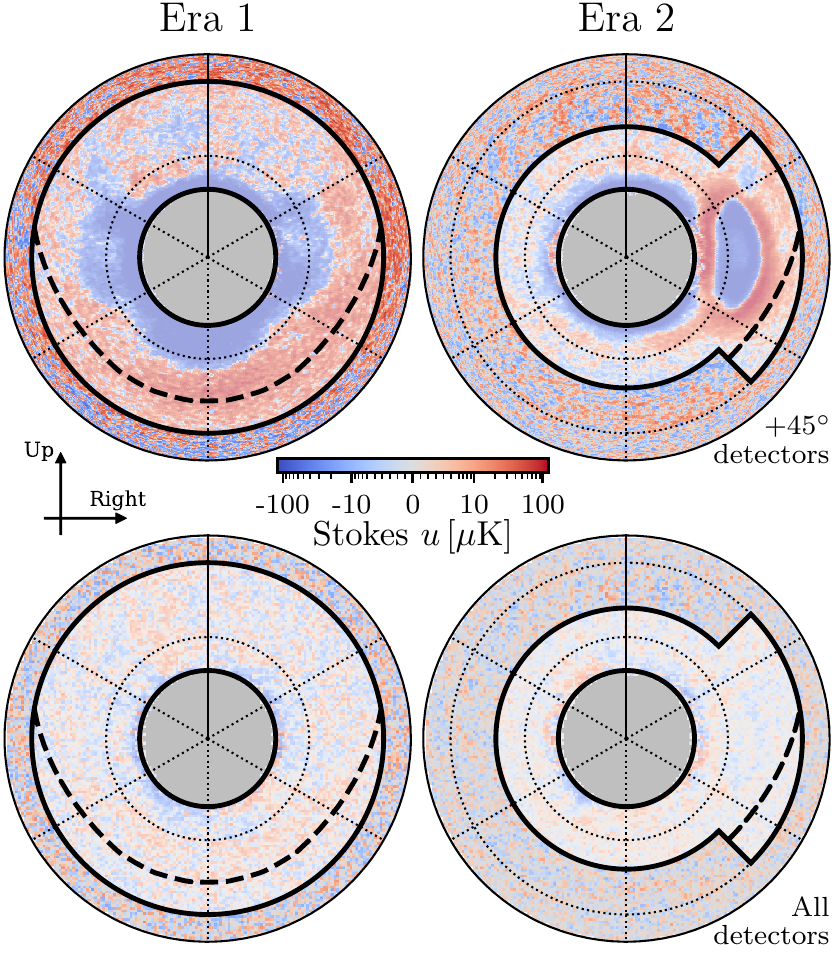}
    \caption{\label{fig:sun-cut}
    The CLASS Sun-centered linear polarization maps and the Sun avoidance cut.
    The maps are made by coadding all $+45^\circ$ oriented detectors (top row) and all detectors (bottom row) in each era in the telescope boresight coordinates (independent of the boresight rotation) and are presented under orthographic projection.
    Some diffuse features around the compact structure and with an opposite sign in amplitude are artifacts due to the baseline adjustment in the mapmaking.
    The location of the lobe on the right in Era 2 corresponds to the location of the opening in the cage of the 90~GHz telescope.
    The features in Sun-centered maps resemble temperature-to-polarization ($T$-to-$P$) leakage that has a canceling effect between $+45^\circ$ and $-45^\circ$ oriented detectors.
    When all detectors are combined, the residual effect is subdominant to the noise.
    The inner black circles show the commanded $20^\circ$ Sun avoidance during the survey; the outer black circles/wedges are the extended boundaries for data flagging based on the relative position of the Sun with respect to the telescope boresight ($60^\circ$ for Era 1 and $40^\circ - 60^\circ$ for Era 2) when the Sun is above the horizon. The position of the horizon changes with the boresight rotation in the telescope boresight coordinates, making regions below the dashed line also accessible from certain boresight angles.
    }
\end{figure}

\paragraph{Source Avoidance} Although the scanning schedule had a $20^\circ$ boresight avoidance from the Sun, no such strategy was applied for the Moon and planets. 
With the $1.5^\circ$ beamwidth at 40~GHz, we flagged all detector data when the Moon is within $3^\circ$ of any detector's pointing direction to prevent the impact of the moonlight through detector crosstalk. Despite the Sun avoidance incorporated into the telescope scan, we were motivated to remove additional data when the Sun was above the horizon due to pickup observed in all detectors at the $-70$~dB level ($100\,\mathrm{\mu K}$ for the Sun) when the Sun encroached on the telescope boresight. The spurious signal in every detector appeared in the same place relative to the telescope boresight position, independent of the pointing of the individual detectors across the $20^\circ$ field of view. This suggests that the issue was not due to direct pickup of sunlight \citep[as is the case with far-sidelobes;][]{xu20}, but another indirect systematic effect, such as a change in the VPM-synchronous signal \citep[VSS;][]{harrington21} due to the exposure of the VPM to sunlight. 
The spurious signal and the corresponding data cuts are shown in Figure \ref{fig:sun-cut}.
In Era 1, we flagged all data when the boresight of the telescope was within $60^\circ$ of the Sun. With the redesigned Era-2 forebaffle extension and roof, we were able to decrease this zone of solar exclusion to $40^\circ$. However, a fan-shaped solar keep-out region extending to $60^\circ$ in the direction of the 90~GHz telescope forebaffle opening was still required. The lower plots of Figure \ref{fig:sun-cut} show the undetectable impact of the spurious signal when the polarization measurements of $+45^\circ$ and $-45^\circ$ oriented detectors are combined. 
In this case, the spurious signal modulated at around 10~Hz decreases by the same amount ($100\,\mathrm{\mu K}$) in both $+45^\circ$ and $-45^\circ$ detectors. For  $+45^\circ$ ($-45^\circ$) detectors, this produces negative (positive) spurious ``polarization'' signals that cancel one another upon combination. Because our survey maps incorporate all of the detectors and cancel the spurious solar signal as in the bottom half of Figure \ref{fig:sun-cut}, this avoidance cut represents a conservative measure to ensure data quality.

\paragraph{VSS Amplitude Cuts} The strongest signal at the modulation frequency of $10\,\mathrm{Hz}$ is the VSS, which serves as a good indicator of a detector's optical response. 
An estimator of the strength of the VSS was computed every 10 minutes for each detector across a span.  A detector's data over the entire span were flagged if the detector's median VSS strength estimator was less than five times the standard deviation of the detector's VSS strength estimator over the span (i.e., the VSS strength estimator had a signal-to-noise, S/N, ratio below five).

\paragraph{Conditioning Cuts}
The final mapmaking operates on 10-sweep segments of data for noise modeling (Section \ref{sec:map-making}).
The 10-sweep segments of data of each detector were dropped for analysis if the retained data fraction was below $52\%$ to improve the stability of filtering and noise model estimation.

\subsection{Pointing}\label{ssec:pointing}
During data acquisition, the pointing model was used to convert the commanded position to the encoder positions used by the servo system. As the DataPkgs were read-in to form the span, the telescope pointing was reconstructed from the recorded encoder data by inverting an updated pointing model through simple iterative successive substitution. In this way, the pointing model used for the analysis was generally not the pointing model used to acquire data: it was revised based on additional calibrations. 
It is shown in \cite{xu20} that the long-term stability of the 40~GHz telescope pointing model is better than 1.4 arcminute. For a detailed description of the pointing model, see Appendix~\ref{sec:pointing-model}.
The computation of detector pointings from the pointing model used an adapted version of \texttt{qpoint}.\footnote{\url{https://pypi.org/project/qpoint}}

\subsection{Calibration}
\label{ssec:calibration}
The spans constructed from the DataPkgs were encoded in raw data acquisition units. 
All span data were first calibrated to physical units of power detected at each bolometer by applying TES responsivity gain factors extracted from 
the current–voltage characteristic ($I$-$V$ load curve) of the detector. 
The uncertainty of each CLASS TES responsivity factor is approximately 0.5\% \citep{appel22}.
Previous CMB experiments have also derived their TES detector calibration from $I$-$V$ data \citep[e.g.,][]{dunner13, abs18}.
The power detected at the bolometers was converted to sky thermodynamic temperature taking into account the optical efficiency of each detector,  the optical depth of the atmosphere at the detector elevation pointing, and the CMB power-to-temperature relationship evaluated at the detector frequency bandpass.
The relative and absolute optical efficiencies of each detector were obtained from source observations such as the Moon or Jupiter \citep{appel19, xu20, dahal22}.

\begin{figure}
    \includegraphics[width=\linewidth]{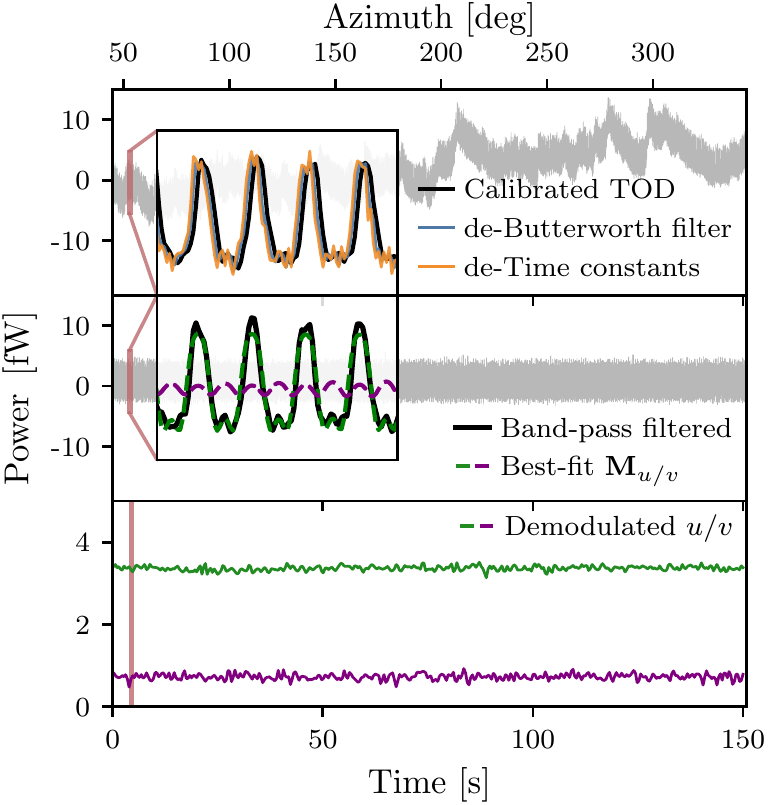}
    \caption{Demonstration of pre-processing and demodulation of a single detector's data over a $150\,\mathrm{s}$ window. 
    \textit{Top:} the raw data calibrated into power units (black) 
    and processed data with the MCE readout Butterworth filtering (blue) and detector time constant (orange) deconvolved.
    The inset plot zooms into a shorter $\sim0.45\,\mathrm{s}$ window (in red) to see the impact over a few VPM modulation cycles. The dominant $10\,\mathrm{Hz}$ component is due to the VSS.
    \textit{Middle:} the deconvolved data were further bandpass filtered around the modulation harmonics lines (black). 
    The VPM transfer functions $\mathbf{M}_{u/v}$ are shown as green and purple dashed lines, 
    with the amplitudes set to the best-fit values from the data segment,  i.e., the demodulated $u/v$ amplitudes. Here the calibration factor $\Delta T_\mathrm{CMB}/\Delta D_{i}$ defined in Equation \ref{eq:temp-calib} is divided out to keep the demodulated data in power units.
    \textit{Bottom:} demodulated $u/v$ time streams. The corresponding range to the zoomed-in region is marked in red.
    The azimuth angle scanned over during this time period is marked on the top axis. 
    The flatness of the demodulated curves indicates the stability of the polarization measurement.
    }
    \label{fig:demodulation-sketch}
\end{figure}

\subsection{Detector Time Constants and Butterworth Filter}
\label{ssec:time-constants-filter}
The sky signal was low-pass filtered twice before being recorded on the disk.
The detectors have time constants that depend on their electro-thermal properties at any given time.
The detector electro-thermal behavior varies in response to changes in optical loading and bias current. 
The detector filtering can be modeled as a single-pole low-pass filter with $3-4\,\mathrm{ms}$ time constant ($f_{3\mathrm{dB}} = 40-50\,\mathrm{Hz}$) applied to the data \citep{appel19,dahal22}.
Furthermore, the MCE sampled the detectors at around $23\,\mathrm{kHz}$ \citep{dahal2020thesis} and applied a fourth-order low-pass Butterworth filter with $50\,\mathrm{Hz}$ cutoff frequency ($3\,\mathrm{dB}$) before the data were downsampled to $201\,\mathrm{Hz}$ and recorded. 

Both of these filters are causal and shift the detector response in time.
They must be deconvolved before further analysis, which requires synchronizing the detector response with the VPM encoder data and detector pointing data.
The first panel of Figure~\ref{fig:demodulation-sketch} illustrates these steps.
The Butterworth filter is a known property of the MCE and was deconvolved as the first step after the data calibration (blue curve). 
Due to the presence of the VSS, the effect of the time-constant convolution shows up as a distinct hysteresis between the VSS and the VPM mirror motion \citep{appel19}.
Taking advantage of this, the time constants were measured by minimizing the hysteresis and were deconvolved from the data as shown by the orange curve.
The detector time constants were found to be mostly stable; therefore, we chose the average value per observation era per detector for the analysis. We assess the potential systematic errors from this choice in Section \ref{ssec:sys-tau}.

\subsection{VPM and Demodulation}\label{ssec:demodulation}
Polarized sky signals as seen by the CLASS detectors are modulated by the front-end VPM; therefore, a demodulation step is needed to recover the sky signal before making maps. 
This section provides an overview of the VPM modulation physics and the demodulation processing.
In-depth descriptions are presented in \cite{chuss06, chus12vpm}, \cite{harrington2018thesis}, and \cite{harrington21}; the summary here highlights the calibration of the VPM transfer function for cosmological analysis.

\subsubsection{Monochromatic VPM Modulation}
\begin{figure}
\includegraphics[width=\linewidth]{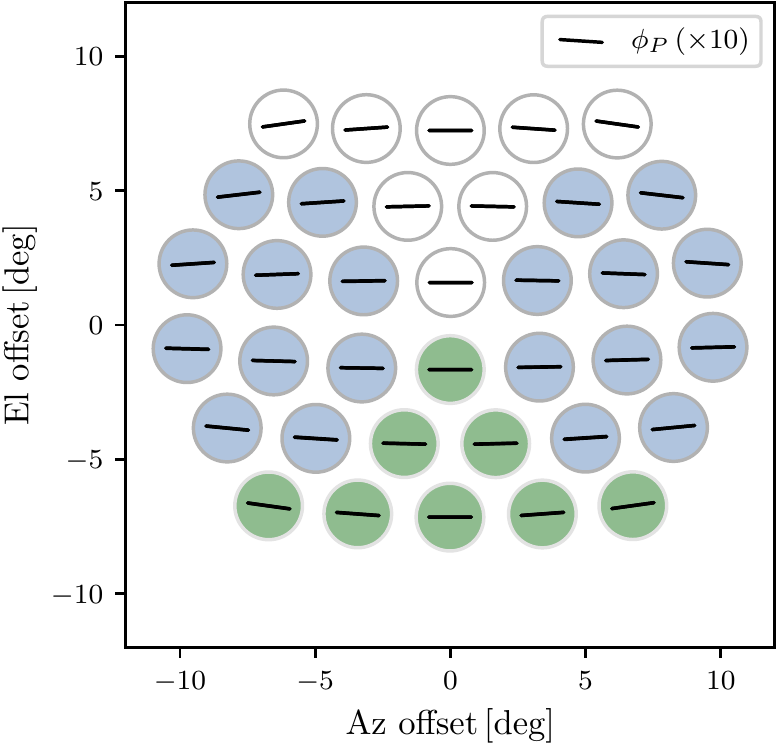}
    \caption{\label{fig:wire-dir}CLASS 40~GHz focal plane layout and the VPM wire grid angle $\phi_P$ as seen by each detector. 
    Each circle represents a pair of orthogonal detectors and is plotted by its pointing offsets on the sky. 
    The orientation of the line within each circle denotes the VPM wire direction as seen by that detector, $\phi_P (\times 10)$, from the geometric modeling; the polarization sensitivity is $45^\circ$ from the wire direction.
    $\phi_P$ varies by $1.6^\circ$ across the focal plane.
    The detectors on the bottom of the plot, (shown in green) see the main polarization calibrator Tau A at the nominal $45^\circ$ elevation scan at all boresight rotations; the blue detector pairs on either side only have partial Tau A coverage depending on the boresight rotation angle, and the unfilled pairs have no Tau A coverage.
    }
\end{figure}

For monochromatic radiation at frequency $\nu$, the intensity reaching each linearly polarized detector, $I_\nu$, as a function of time $t$ and the grid-mirror distance $z(t)$ is:\footnote{The notation slightly deviates from that in \cite{harrington21}. We also omit in 
 notations the implicit time dependency through $z(t)$.}
\begin{align}
    I_\nu(z, t)  = A_z(z)i_\mathrm{VPM}(t)+
    \begin{bmatrix}
        S_i(z) \\S_q(z) \\S_u(z) \\S_v(z)
    \end{bmatrix}^T \,
    \begin{bmatrix}
        i(t)\\q(t)\\u(t)\\v(t)
    \end{bmatrix}
    \label{eq:mono-modulation-function}
\end{align}
where $A_z(z)i_\mathrm{VPM}$ includes the emission from the VPM grid and mirror that makes up the majority of the VSS at 40~GHz; $S_{i,q,u,v}$ terms are \textit{modulation functions}, which we describe below; the lowercase $i,q,u,v$ are used for Stokes $I, Q, U, V$-like signals in the VPM coordinate system, where $u$ is the polarization state that is modulated by the VPM, and $q$ is the orthogonal state.
The linear polarization recorded by each detector after modulation by a common VPM is related to the sky polarizations $[Q, U, V]$ by a rotation angle $\phi_P$ such that:
\begin{align}
    \begin{bmatrix}
        q\\
        u\\
        v 
    \end{bmatrix}=
    \begin{bmatrix}
        \cos2(\phi_P+\gamma) & \sin2(\phi_P+\gamma) & 
        0\\
        -\sin2(\phi_P+\gamma) & \cos2(\phi_P+\gamma) &
        0\\
        0 & 0 & 1
    \end{bmatrix}
    \begin{bmatrix}
        Q\\
        U\\
        V
    \end{bmatrix},
    \label{eq:wire-angle}
\end{align}
where $\gamma$ is the detector position angle on the sky, and $\phi_P$ corresponds to the VPM wire direction as seen from each detector (Figure \ref{fig:wire-dir}), both of which are functions of the detector pointing offset from the center of the focal plane. 
The throw $z(t)$ is changing at about $10\,\mathrm{Hz}$ so that the polarization modulation functions $S_{u/v}$ move the signal in $u(t)$ and $v(t)$ to higher frequencies (around the harmonic series of the VPM fundamental frequency), away from the slowly varying noise component in the unpolarized term $i(t)$ \citep{tatarski61,church95, morris2021}. The throw range is chosen to optimize the linear polarization observation so that $S_{u/v}$ is mainly nonzero around $10/20\,\mathrm{Hz}$.
For an ideal VPM with zero emissivity, $A_z=0$ and $i$, $q$ are not modulated ($S_i=1$ and $S_q$ is constant).
The microwave frequency dependence of the Stokes parameters and the modulation functions in Equation \ref{eq:mono-modulation-function} should be implicitly assumed.
We refer the reader to \cite{harrington2018thesis} for detailed derivations of the modulation functions $S_{u,v}$.

\subsubsection{VPM Transfer Functions}\label{sssec:vpm-transfer-function}
The modulated power $D$  detected by the TES bolometers is the bandpass and beam integrated version of Equation \ref{eq:mono-modulation-function}:
\begin{align}\label{eq:flux-integral}
    D&= \eta \int \mathrm{d}\Omega \int A_e(\nu) B_\nu(\mathbf{\Omega})f_\nu I_\nu\,\mathrm{d}\nu \notag \\
    &= \int \eta\frac{\lambda^2}{2} f_\nu I_\nu \mathrm{d}\nu,
\end{align}
where $\eta$ is the optical efficiency that accounts for all loss terms in the optics, $B_\nu(\Omega)$ and $A_e(\nu)$ are the beam profile and the effective telescope aperture area, $f_\nu$ is the detector bandpass, and $\lambda$ is the wavelength.
The identity \citep{Pawsey&Bracewell1954}
\begin{equation}
    \int A_e B_\nu(\mathbf{\Omega}) \mathrm{d}\,\Omega=\frac{\lambda^2}{2}
\end{equation}
is assumed for a single-mode detector observing a beam-filling source (e.g., CMB and extended foreground emission).

For unpolarized radiation, the received power is calibrated to the thermodynamic temperature unit by assuming a blackbody spectrum $I_\mathrm{BB}$:
\begin{align}
    \Delta D_i &= \Delta T_\mathrm{CMB} \int \eta\frac{\lambda^2}{2} f_\nu S_i \frac{\partial \I_\mathrm{BB}(\nu)}{\partial T_\mathrm{CMB}} \mathrm{d}\nu \\
    & = \Delta T_\mathrm{CMB}\int \eta k_\mathrm{B} f_\nu \frac{x^2 e^x}{(e^x-1)^2} \mathrm{d}\nu. \label{eq:temp-calib}
\end{align}
Here, we use $\Delta$ to denote the spatial fluctuation of the quantity; $x\equiv h\nu/k_\mathrm{B}T_\mathrm{CMB}$ where $h$ and $k_\mathrm{B}$ are the Planck and Boltzmann constants, and $T_\mathrm{CMB}$ is the CMB temperature \citep{mather94}. 
The coefficient $\Delta T_\mathrm{CMB}/\Delta D_{i}$ is the intensity calibration factor obtained from the Moon observations \citep{appel19,dahal22}.

For polarization modulated by a VPM, Equation \ref{eq:flux-integral} can be formally cast into a matrix product
\begin{align}
    \Delta D = 
    \begin{bmatrix}
        \tilde{M}_u\,\,\tilde{M}_v
    \end{bmatrix} \,
    \begin{bmatrix}
        \Delta T_{u, \mathrm{s}} \\
        \Delta T_{v, \mathrm{s}}
    \end{bmatrix},
\end{align}
where
\begin{equation}
    \tilde{M}_{u/v}(z) = \int \eta k_\mathrm{B} f_\nu S_{u/v}(z,\nu) s_{u/v}(\nu)\,\mathrm{d}\nu.
\end{equation}
Here we have ignored the unmodulated linear polarization component $q(t)$ and substituted the polarization terms $u(t)$ and $v(t)$ from Equation \ref{eq:mono-modulation-function} into Equation \ref{eq:flux-integral}. 
The polarization intensity $u(t)/v(t)$ are separated as the product of their effective temperature $T_{u/v,\mathrm{s}}$ evaluated at an arbitrary reference frequency, and the associated spectral shapes $s_{u/v}(\nu)$ such that $u(\nu)\equiv2 k_\mathrm{B}T_{u,\mathrm{s}}s_{u}(\nu)\lambda^{-2}$ (likewise for $v(\nu)$).
Note that the effective temperature depends on the spectrum of the source, which may be different for linear and circular polarization and might vary across the sky.
For cosmological analysis, the source effective temperature is calibrated to thermodynamic temperature $T_{u/v}$\footnote{The subscript `$\mathrm{CMB}$' is omitted for quantities in thermodynamic units unless noted otherwise.} through
\begin{equation}
    \frac{\Delta T_{u/v}}{\Delta T_{u/v, \mathrm{s}}} = 
    \frac{\int f_\nu s_{u/v}\,\mathrm{d}\nu}{\int f_\nu \frac{x^2 e^x}{(e^x-1)^2}\,\mathrm{d}\nu}.
\end{equation}
Therefore, the \textit{calibrated transfer function}
\begin{align}
    M_{u/v} &= \left(\frac{\Delta T_{u,v}}{\Delta T_{u/v,\mathrm{s}}}\right)^{-1}\tilde{M}_{u/v}\\
    &= \left(\frac{\Delta T_\mathrm{CMB}}{\Delta D_{i}}\right)^{-1}\frac{\int f_\nu S_{u/v}(\nu) s_{u/v}(\nu)\,\mathrm{d}\nu}{\int f_\nu s_{u/v}(\nu)\,\mathrm{d}\nu},
\end{align}
relates the modulated power to polarization intensity in thermodynamic units, where the first term corresponds to the power-to-thermodynamic temperature unit conversion factor for intensity, and the second term is the uncalibrated transfer function.

\subsubsection{Demodulation}\label{sssec:demodulation}
The modulated time stream calibrated in thermodynamic temperature units is\footnote{It should be assumed that a CLASS bolometer only measures the relative power fluctuation, so we drop the $\Delta$ notation in the text below for simplicity.}
\begin{align}
    \label{eq:modulation-equation}
    \mathbf{D} = 
    \begin{bmatrix}
        \mathbf{M}_u\,\,\mathbf{M}_v
    \end{bmatrix} \,
    \begin{bmatrix}
        \mathbf{T}_u\\
        \mathbf{T}_v
    \end{bmatrix},
\end{align}
where the bold vector symbol is used to emphasize that each of the quantities is a time stream of length $n_\mathrm{samp}$.
The transfer function time streams $\mathbf{M}_{u/v}$ are evaluated for the VPM grid-mirror distance encoder data ($z(t)$, synchronously sampled with the data).
The goal of demodulation is to solve the sky polarization signal $\mathbf{T}_{u/v}$ from the raw time stream $\mathbf{D}$.
The least-squares solution to Equation~\ref{eq:modulation-equation} is \citep[see also,][]{harrington21}
\begin{align}
    \
    \begin{bmatrix}
           \mathbf{T}_u \\
           \mathbf{T}_v
    \end{bmatrix} 
    &= \mathfrak{L}\left[\mathbf{M}^T\mathbf{M}\right]^{-1} \mathfrak{L}\left[ \mathbf{M}^T\mathbf{D}\right],
\label{eq:demod_solution}
\end{align}
where $\mathbf{M} = [\mathbf{M}_u\, \mathbf{M}_v]$ is the matrix form of the transfer function. 
Prior to the least-squares solving, a bandpass filter was applied that includes the first five harmonics of the VPM frequency ($\sim 10-50\,\mathrm{Hz}$) with a margin of $f_c=0.5\,\mathrm{Hz}$ for the $1\,\mathrm{deg}\,\mathrm{s}^{-1}$ az scan and $f_c=1.0\,\mathrm{Hz}$ for the $2\,\mathrm{deg}\,\mathrm{s}^{-1}$ scan, as informed by the beam-crossing timescale of the 40~GHz telescope.
The bandpass filter was applied to both $\mathbf{M}$ and $\mathbf{D}$ so that the effect of this filter does not bias the solution. 
The solution was then filtered with an antialiasing low-pass filter ($\mathfrak{L}$) with cutoff frequency $f_c$ before downsampling to data rates of $1.45\,\mathrm{Hz}$ for the $1\,\mathrm{deg}\,\mathrm{s}^{-1}$ az scan and $2.42\,\mathrm{Hz}$ for the $2\,\mathrm{deg}\,\mathrm{s}^{-1}$ scan. Unlike the bandpass filter, which is accounted for in the demodulation least-squares solution, the low-pass filter in principle can remove signal as it is not accounted for in the mapmaking step. In practice, however,
the cutoff frequency $f_c$ was chosen to have minimal suppression of the signal beyond beam smoothing; its residual impact on the mapping transfer function is characterized in Section \ref{ssec:map-transfer-function}.
The middle panel of Figure \ref{fig:demodulation-sketch} visualizes this process. 
The least-squares fit finds the best solution that matches the amplitude of the $u/v$ transfer functions, which are interpreted as the sky polarization intensity through Equation \ref{eq:wire-angle}.

A preliminary gap-filling was performed after the demodulation to fix the discontinuity of the data due to data selection (Section \ref{ssec:tod-selection}) and splitting of data for consistency tests (Section \ref{ssec:null-result}).
The demodulated data were separated into low- and high-passed components by a rolling top-hat kernel of 50 samples in width.
The gaps in the demodulated data were filled with the linear interpolation of the low-pass filtered component and then injected with white noise sampled from the white noise estimation of the high-pass filtered component.
This treatment ensures the basic continuity of the data for subsequent operations but does not necessarily preserve the low-frequency noise structure. 
A dedicated gap-filling for this purpose is introduced in Section \ref{ssec:gap-filling}.

\subsubsection{VPM Parameters}\label{ssec:vpm-parameters}
Calibration of the VPM parameters is essential to the recovery of the polarization signals.
The parameters used for cosmological analysis were determined through a minimization process of mixing between the linear and circular polarizations (polarization leakage). 
The model parameters included the incident angle of light onto the VPM (per VPM grid period), an overall offset in the grid-mirror distance (per grid period), an overall shift in the bandpass centroid (per Era), the spectral index of the linear polarization, and an additional power-law correction to the atmospheric circular polarization spectrum to account for model uncertainties \citep{petroff20}. 

For each set of parameters, the demodulated data $u/v$ were swapped such that the linear polarization stream $u$ was mapped into horizontal coordinates as an intensity-like signal (to probe for leakage from atmospheric circular polarization into $u$) and the circular polarization stream $v$ was mapped into equatorial coordinates as linear polarization signals (to probe for leakage from Galactic linear polarization into $v$). 
We defined the following polarization-leakage statistic:
\begin{equation}
\label{eq:grid-chi2}
\chi^2 = \sum_{j=1}^{3} \left(\frac{A_{j,u}}{\sigma_{j,u}}\right)^2+\left(\frac{A_{j,v,\mathrm{bot}}}{\sigma_{j,v,\mathrm{bot}}}\right)^2+\left(\frac{A_{j,v,\mathrm{top}}}{\sigma_{j,v,\mathrm{top}}}\right)^2,
\end{equation}
where for each of the three VPM grids labeled by $j$, $A_{u}$ is the amplitude of the atmospheric circular polarization model \citep[][see their Figure 3]{petroff20} fit to the horizontal-coordinate maps created from linear polarization ($u$), and $A_v$ is the correlation of the WMAP $Q$-band linear polarization maps around the Galactic plane with the ``linear polarization'' equatorial-coordinate maps generated from circular polarization ($v$).
For linear polarization, the (quasi-)azimuth-synchronous signals (Section \ref{ssec:map-filter}) were not filtered and left systematics in the horizontal coordinates; therefore, the azimuth range $-10^\circ$ to $110^\circ$ was avoided when computing the circular polarization model amplitude $A_u$.
For the circular polarization, maps were made separately for detectors on the top/bottom of the focal plane (see $A_{v,\mathrm{top}}$ versus $A_{v,\mathrm{bot}}$) to better break the degeneracy between the tilt of the VPM and the grid-mirror distance. 
The uncertainties of these leakage amplitudes $\sigma$ were evaluated with noise-only simulations.

The $\chi^2$ values over the entire VPM parameter space were first sparsely explored with 250 Latin hypercube samples \citep{mckay2000LHS}.
Demodulated data and maps were made for each of the samples, and the $\chi^2$ values were computed according to Equation \ref{eq:grid-chi2}.
Another 250 samples were drawn near the (approximate) minimum of $\chi^2$ and evaluated in the same way.
With a total of 500 evaluations, the rest of the parameter space was parameterized with Gaussian process regression. 
Finally, we used the Markov Chain Monte Carlo method to determine the best-fit parameters that globally minimize the $\chi^2$ values.

For the cosmology maps, we adopted the instrumental parameters and the circular polarization spectrum correction from the minimization process above but used the linear polarization spectrum from \cite{spass18} with a spectral index of $-0.7$ because the minimization process focuses on the Galactic region where the linear polarization index could be different from the rest part of the sky due to the variation of the synchrotron index \citep[predominately from synchrotron;][]{gold09fore,fuskeland14,spass18,planck18IV,quijote23} and the mixture with the CMB. 
The impact of the uncertainties of the linear polarization spectral index and the other parameters from the leakage minimization are further characterized in Section \ref{ssec:sys-vpm-tf}.

\section{Mapmaking}\label{sec:map-making}
The raw CLASS data can be formally modeled as
\begin{align}
    \mathbf{D}(t) = 
    \mathbf{M}(t)\,\mathbf{P}(t)\,m + \mathbf{n},
     \label{eq:data-model}
\end{align}
where $\mathbf{M}(t)$ is the modulation transfer function introduced in Equation \ref{eq:modulation-equation}, $\mathbf{P}(t)$ is the pointing matrix that transforms polarization Stokes parameters from the sky coordinates to the VPM coordinates (i.e., the matrix in Equation \ref{eq:wire-angle}), $m=[Q, U, V]$ are the sky polarization maps, and $\mathbf{n}$ represents the raw data noise. 
The demodulation process described in the previous section partially solves this equation and yields intermediate demodulated data
\begin{align}
    \mathbf{d}(t) &\equiv 
    \begin{bmatrix}
        \mathbf{T}_u \\
        \mathbf{T}_v
    \end{bmatrix} 
     = \mathbf{P}(t)\cdot m + \begin{bmatrix}
        n_u(t) \\
        n_v(t)
    \end{bmatrix}, \label{eq:map-making-model}
\end{align}
where $n_u$ and $n_v$ are noise in the demodulated data.
This section describes the process of solving the polarization maps from $\mathbf{d}(t)$.
We start with filtering the demodulated data to reduce systematics not accounted for by the data model. We then describe the noise model and gap-filling methods for the demodulated time streams and how they are applied for maximum-likelihood mapmaking.
Finally, we present the maps and the associated transfer functions due to the filtering.
The CLASS mapmaking algorithms are developed based upon the public code \texttt{minkasi}\footnote{\url{https://github.com/sievers/minkasi}} \citep{romero20}.

\subsection{Filtering}\label{ssec:map-filter}
The modulation technique and the demodulation process produce polarization time streams that are mostly immune to atmospheric fluctuations and intensity-like systematics from the sky and the ground.
However, these data were found to have systematic signals that may be traced back to polarized environmental emission, $T$-to-$P$ leakage, electronic pickup from the instrument, etc. 
Time-domain filters were designed for each of the cases and were jointly fit and removed from the polarization time streams as
\begin{equation}
    \mathbf{d} \Rightarrow \mathbf{d} - \mathbf{F}(\mathbf{F}^\mathrm{T}\mathcal{M}\mathbf{F})^{-1}\mathbf{F}^\mathrm{T}\mathcal{M}\mathbf{d},
\end{equation}
where $\mathbf{F}$ is a collection of column vectors that include all systematic signal models, which we describe in the rest of this section; $\mathcal{M}$ is a time-domain mask to prevent biasing the filter. The mask comprises the TOD selection mask (Section~\ref{ssec:tod-selection}) and a linear polarization mask that vetoes the brightest $3.6\%$ of the sky in synchrotron polarization \citep{planck18IV} around the Galactic plane.
Nevertheless, this approach filtered out sky modes that mimic systematics in the time domain, especially at large angular scales. 
The impact is quantified by the mapping transfer functions in Section \ref{ssec:map-transfer-function}.

\subsubsection{Azimuth Servo Motor Signal}
A spurious signal was found to be synchronous with the az-servo motor current (thus, with the az-velocity of the telescope), having a peak-to-peak amplitude up to several times the VSS amplitude ($\sim 5\,\mathrm{fW}$).
A set of harmonic components was filtered to remove this signal:
\begin{equation}
    \mathbf{F}_\mathrm{azvel} =\Bigl\{\exp\bigl[im\phi_\mathrm{8\pi}(t)\bigr]\Bigr\}_{m=1}^{3}\footnote{The bracket with sub/superscripts indicates a set of filter basis and its parameter range.}.
\end{equation}
Here, $\phi_{8\pi}$ is the azimuth of the telescope with the $8\pi$ period (which accounts for both the positive and negative azimuth velocity sweeps). 
This component was filtered from the linear and circular polarization data every 3 hr.

\subsubsection{Wind-induced Signal}\label{ssec:wind-signal}
A quasi-azimuth-synchronous signal is present in the demodulated linear polarization data that correlates with the wind recorded by the WeatherHawk\footnote{\href{https://web.archive.org/web/20150307230734/http://www.weatherhawk.com/s232dc}{http://www.weatherhawk.com/s232dc}} weather station installed close to the CLASS telescope.
The weather station provides wind speed information through a cup anemometer (starting threshold at $0.78\,\mathrm{m\,s^{-1}}$) and wind direction through a vane.

\begin{figure}
\includegraphics[width=\linewidth, 
trim={0 0.\linewidth, 0 0.\linewidth},clip]{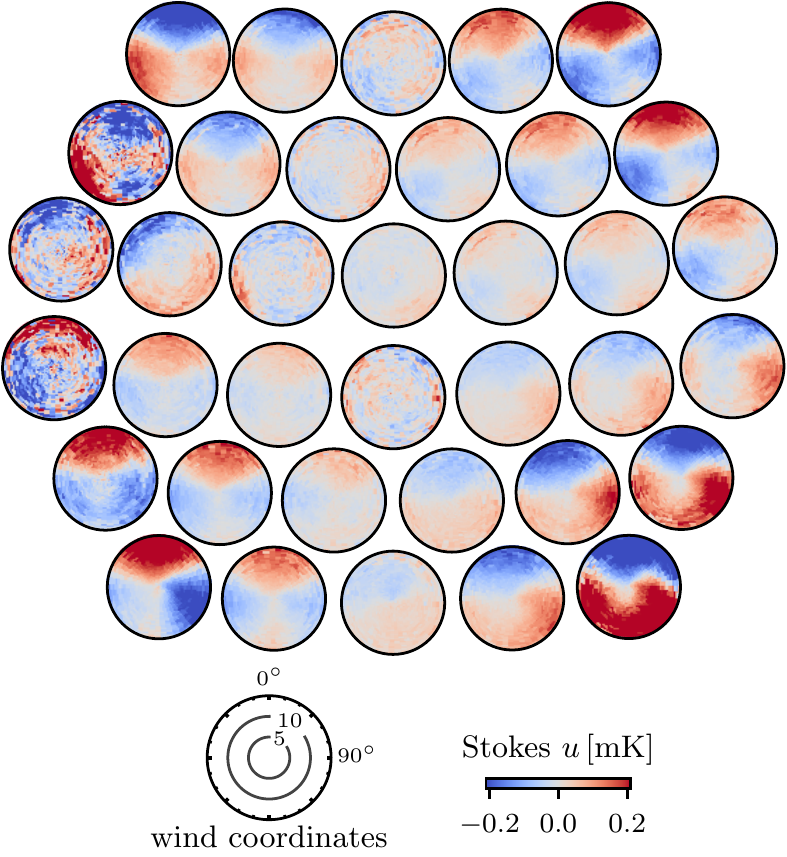}
\caption{
The wind-induced signal across the focal plane when the plastic closeout was on and the forebaffle extension was blackened.
Each panel shows the demodulated linear polarization signal $u$ binned in the wind bearing angle and wind speed coordinates for the $-45^\circ$ oriented detector in each feedhorn; the other detector in the pair shows a similar signal.
As indicated by the legend, the azimuth angle of the polar plots corresponds to the wind bearing angle, and the radial axis marks the wind speed in units of meters per second.
The panels are laid out by the detector pointing offsets, similar to Figure \ref{fig:wire-dir}, with the detectors pointing at higher elevations at the top.
Within each panel, the wind-related systematics are mostly confined in the quadrant when the telescope is facing toward the wind. 
The signal amplitude varies across the focal plane with a quadrupole pattern that increases from the center toward the edge of the focal plane and also rises with the wind speed.
The whirlpool-shaped structure in some of the edge detectors is due to the ground pickup (Section \ref{ssec:az-signal})
}
\label{fig:2d-wind-template}
\end{figure}

\begin{figure}
\includegraphics[width=\linewidth, trim={0, 0.17\linewidth, 0.01\linewidth, 0.1\linewidth}, clip]{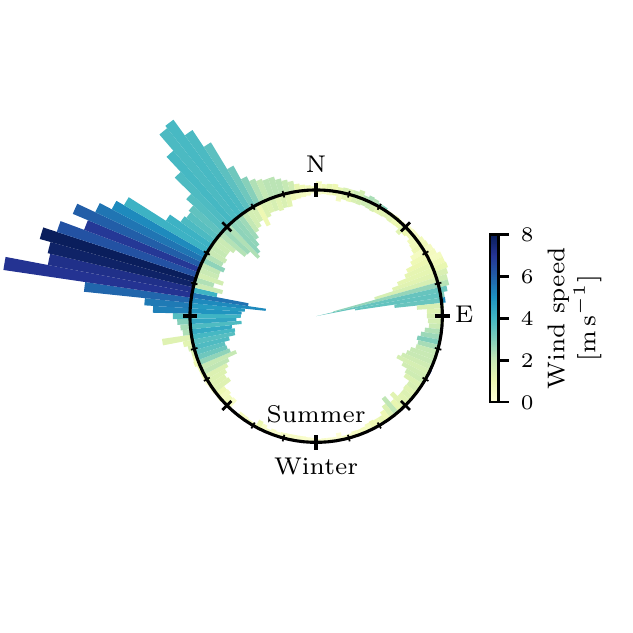}
\caption{
    Wind direction distribution at the Atacama site as measured by the CLASS weather station over the 6 yr survey. 
    The wind direction distribution is shown as the histogram in the outer and inner ring for the Austral winter (June to August) and summer (December to February), respectively.
    The histograms are color-coded by the mean wind speed in each bin.
}
\label{fig:wind-stats}
\end{figure}

Figure~\ref{fig:2d-wind-template} shows the demodulated linear polarization signal for each of the $-45^\circ$ oriented detectors as a function of the bearing angle of the wind with respect to the telescope azimuth pointing and the wind speed (the radial axis).
The quadrupole feature across the focal plane that peaks around $0^\circ$ and at high wind speed is due to the deformation of the plastic closeout film at the telescope's optical entrance when pressed by the wind.
The blue/red features toward the south are due to arbitrary baseline adjustments for this plot. 
The wind signal was found to be consistent at different boresight configurations and scales roughly linearly with the wind speed. 
As shown in Figure \ref{fig:wind-stats}, the prevailing wind at the site came from the northwest during the austral winter and had significant contributions from the east in summer, with a slight shifting in direction throughout the night (not shown in the plot); therefore, this wind-induced signal left a systematic error that is covariant with azimuth pointing and thus with sky signals.

Instrumental mitigation of this issue started in September 2021 by removing the covering plastic during observations.
For the time period with the closeout on, the filter components took the form:
\begin{equation}
\mathbf{F}_\mathrm{wind} = \Bigl\{\gamma_\mathrm{w}\exp\bigl[im(\phi_\mathrm{w}-\phi(t))\bigr]\Bigr\}_{m=1}^{5},
\end{equation}
where $\gamma_\mathrm{w}$ and $\phi_\mathrm{w}$ are the wind speed and wind direction measured by the weather station, $\phi(t)$ is the azimuth pointing of the telescope with $2\pi$ period.
Wind data were up-sampled from the original rate at $0.5$ Hz to align with the demodulated data.
This filter was applied to the linear polarization time streams every 2 hr.

\subsubsection{Azimuth-synchronous Signal}\label{ssec:az-signal}
\begin{figure*}
    \begin{center}
    \includegraphics[width=1\linewidth]{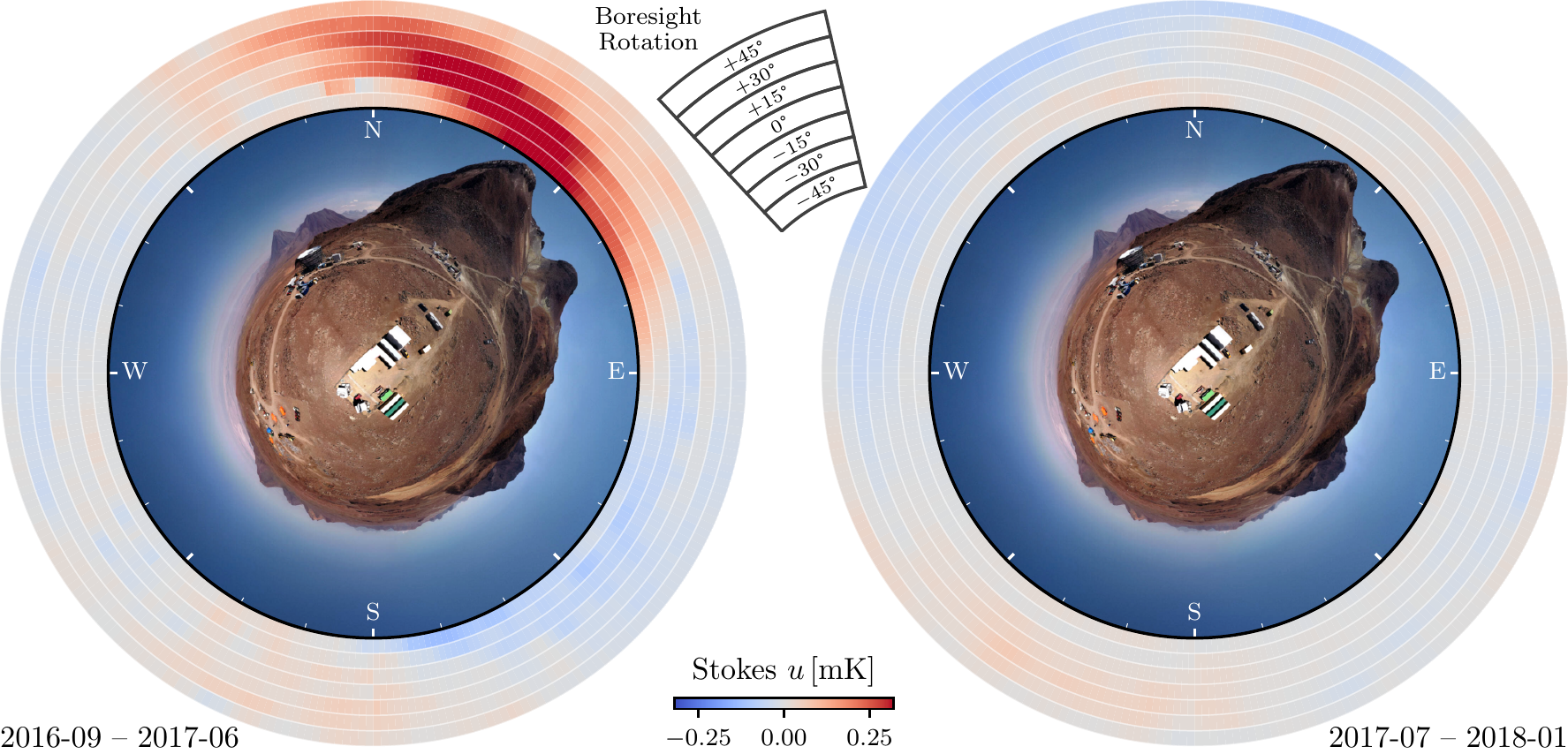}
    \end{center}
    \caption{
        \label{fig:panorama}
        Ground pickup signals and the landscape around the CLASS site.
        The concentric annuli show the demodulated linear polarization signal ($u$) for a pair of detectors at the bottom edge of the focal plane (linear polarizations after demodulation have the same sign between the pair and are averaged to enhance the signal) binned in the telescope azimuth angles for the seven boresight rotations. 
        {\it Left:} data taken prior to 2017-06 when the circular baffle was not blackened. Of note is the Cerro Toco mountain peak ($15^\circ$ elevation) at around $45^{\circ}$ in azimuth, which aligns with the peak of the $u$ signal. 
        For this edge detector pair, the signal from the mountain appears earlier or later in the telescope azimuth coordinates depending on the boresight rotations. 
        {\it Right:} data from 2017-07 to 2018-01, where the interior of the baffle extension was blackened, and before the replacement with the asymmetric double baffle extension and new baffle roof. The Era~2 configuration gave a reduction in ground pickup comparable to the right panel in all but the outermost detectors, regardless of the blackening state.
     }
    
\end{figure*}

Several systematic issues were found to contribute to an azimuth-synchronous signal.
The metal surface of the telescope cage reflected ground emission and produced signals in both intensity and linear polarization. 
This was mitigated sequentially by blackening the interior of the cage and the forebaffle with microwave absorbers. 
At the beginning of Era 2, the circular forebaffle extension was replaced by the double baffle, which has an asymmetric shape and a larger opening angle to accommodate the new 90~GHz telescope. 
The forebaffle extension blackening was removed at the beginning of 2019.
Figure~\ref{fig:panorama} presents an example of this signal in the linear polarization stream for a pair of detectors in Era~1 before and after the blackening.\footnote{In Era~2, regardless of blackening, the ground pickup was reduced in amplitude comparable to the Era~1 blackened state for all but the outer detectors, likely due to the enlarged opening angle of the double baffle and new baffle roof design.}
The linear polarization signal at each boresight angle is binned in the telescope azimuth coordinates as seven separate rings. 
The reflection picked up terrestrial emission that correlated with the Cerro Toco mountain toward the northeast and depended on the boresight rotation of the telescope (which changed both the polarization angle and the pointing elevation of the detector).
The atmospheric circular polarization due to the Zeeman splitting of molecular oxygen magnetic dipole transitions is defined by the azimuth angle from the magnetic North and the pointing elevation \citep{petroff20}. 
This is a smooth az-synchronous systematic for sky circular polarization and was also a potential bias to the linear polarization measurement through leakage from imperfect modeling of the VPM transfer function (Section \ref{sssec:demodulation}).
The aforementioned wind signal may also have left a residual in the azimuth due to inaccuracy in the wind data and modeling errors.

In addition to these relatively stable components, it was found that the electronic coupling to the detector caused an azimuth-synchronous signal that varies on time scales of a few hours. 
This is likely related to the wiring of the detectors since the detectors on the right-hand side of the focal plane, which connect to shorter cryogenic wires, show more stable signals at lower amplitudes. 
This electronic pickup was also notably improved since the deployment of Era~2. 

A set of harmonic filter components was employed to mitigate these signals:
\begin{equation}
\mathbf{F}_\mathrm{az} = \Bigl\{ \exp\bigl[im\phi(t)\bigr]\Bigr\}_{m=1}^{15/10},
\end{equation}
where the $u/v$ time streams were fit with 15/10 harmonic components, respectively.
The filters for circular polarization were evaluated every 3 (4) hr for Era~ 1(2); for linear polarization, the timescale was chosen to be 2 hr for the left detectors, and 3 (6) hr for Era~1(2) for the right detectors.
The amplitude of each of the components in $\mathbf{F}_\mathrm{az}$ was determined separately for the positive and negative az-velocity regions to further mitigate the difference in the az-synchronous signals that were correlated with the az-servo motor. 
Of all of the filters described in this section, these removed the most celestial signal.

\subsubsection{Camera RFI}\label{ssec:camera-signal}
Beginning on 2017 August 9 through 2018 June, the 40~GHz detectors experienced RFI pickup from a camera installed inside the mount cage.  
The RFI (and its harmonics) around the VPM modulation frequencies produced a slowly 
varying harmonic structure in the demodulated polarization time streams.
Since linear and circular polarization were predominantly modulated 
by the first and second harmonics of the VPM modulation frequency,
the contamination in the demodulated data was mostly confined around $0.51$ and $1.53\,\mathrm{mHz}$ for linear polarization, and around $1.02\,\mathrm{mHz}$ for circular polarization.

For the time period with cage cameras on during observations (Section \ref{ssec:changes}), a set of harmonic lines at these frequencies was fit and removed from the demodulated data every 3 hr. 
This is a gentle filter to the sky signal as the frequencies are below the scanning frequency at $2.7\,\mathrm{mHz}$ (for $1\,\mathrm{deg\,s}^{-1}$ scan).

\subsection{Noise Model}\label{ssec:N-mod}
Like other ground-based CMB experiments, CLASS (demodulated) data are noise dominated; therefore, the noise term in Equation \ref{eq:map-making-model} needs to be carefully modeled to achieve optimal sensitivity in the maps.

The demodulated noise $n_u$ and $n_v$ were similarly white at high frequencies and correlated over long time scales \citep[$f_{\mathrm{knee}}, \approx5-20\,\mathrm{mHz}$,][]{harrington21, cleary22} but were also distinct due to the difference in the modulation functions and the nature of linear and circular polarization. 
In particular, atmospheric signals sourced long-time-scale correlated noise across the focal plane \citep{tatarski61,church95,wollack97,lay&halverson00,errard15,morris2021}, some of which were linearly polarized \citep{Takakura2019}, or could have impacted linear polarization through $T$-to-$P$ leakage. 
The emission from the VPM was predominantly covariant with the linear polarization signal, and its slow temporal variation was another potential source of long-term instability \citep{miller16, harrington21}.
The electronics in the readout system \citep{reintsema03,nist_tdm_mux13b} could also have contributed to the correlated noise in the demodulated data, which was mainly manifest as common correlation features between pairs of detectors and, to a lesser extent, within each readout column \citep{dunner13, harrington21}.
Due to the covariance of the modulation transfer functions, $n_u$ and $n_v$ were also expected to be correlated at all scales.

Formally, the noise model takes the following form in Fourier space:
\begin{equation}
    \mathbf{N}(f) = \mathbf{V}^{\dagger}\mathbf{N}_M(f) \mathbf{V} + \mathbf{N}_D(f),
\end{equation}
where $\mathbf{N}$ is the covariance matrix among all detectors and between linear and circular polarizations, $\mathbf{N}_M$ contains the power spectra of the common modes that are projected to each detector through $\mathbf{V}$, and $\mathbf{N}_D$ is a collection of power spectra per feedhorn.
The construction of the common modes $\mathbf{N}_M$ is informed by the noise properties of the data. 
Following \cite{dunner13}, a singular value decomposition (SVD) was performed on the low-frequency (below $0.1\,\mathrm{Hz}$) part of the data to identify modes with dominant singular values (above 3.5 times the median); subsequently, a second SVD was applied on the full frequency range to further find common modes after the removal of the low-frequency modes. 
An SVD per readout column was then used to search for residual correlated modes unique to each readout column. 
There are eight columns in the 40~GHz telescope focal plane, each containing eight or ten optical detectors. 
This hierarchical construction typically found $\sim20$ modes among the 144 detector-Stokes time streams in total.
The per-feedhorn component $\mathbf{N}_D$ captured the rest of the noise power as a block-diagonal matrix with four-by-four blocks that describe the covariance between the two polarization states of the two paired detectors associated with the same feedhorn.
This noise model was estimated for all the data over every 10 sweeps, which is about 2 hr for $1\,\mathrm{deg\,s}^{-1}$ scan.
To facilitate fast evaluations of the noise model and its inversion, the spectra were logarithmically (linearly) binned above (below) $50\,\mathrm{mHz}$. Despite the filtering described in Section \ref{ssec:map-filter}, the azimuth-synchronous systematics were not completely removed as the shapes of the ground pickup, wind-induced signals, and electronic coupling gradually vary over time, and contribute to the noise model.
The binned power spectrum has the advantage of capturing this residual power and allows for the down weighting of the data at around the scanning frequencies.
The matrix construction above ensures a low-rank $\mathbf{N}_M$ and trivially invertible $\mathbf{N}_D$, allowing for an efficient inversion using the Woodbury identity \citep{woodbury1950}.

The noise model was used to optimally weight the data for mapmaking and can be directly sampled to create noise simulations.
However, the model estimated above can be biased due to (1) the missing data in the time streams from data selection, and (2) the direct estimation of the covariance matrix from data that contains both noise and signal. 
The second type of bias would further bias the sky signal estimation when used in Equation \ref{eq:map-making-model}. 
These issues are addressed, respectively, with iterative methods in Sections \ref{ssec:gap-filling} and \ref{ssec:ML-map-making}.

\subsection{Gap-filling}\label{ssec:gap-filling}
As mentioned in Section \ref{sssec:demodulation}, the preliminary gap-filling does not preserve the low-frequency noise properties of the data within the gap, and the noise model directly estimated from this may be biased.
To improve this, we express the mask-aware data likelihood as
\begin{align}    
      -2\ln\mathcal{L}(\tilde{d}\mid d) &= \tilde{d}^T\mathbf{N}^{-1}\tilde{d} + (d-\tilde{d})^T\mathbf{N}^{-1}_\infty(d-\tilde{d})\\
&= -2\ln \int_t \mathcal{P}(\tilde{d}, t| d) \mathrm{d}\,t,
\end{align}
by introducing the better gap-filled data $\tilde{d}$. 
Here, $\mathbf{N}$ is the noise model estimated from the preliminary gap-filled data or from a previous iteration, and $\mathbf{N}_\infty$ is the noise model for the masked data that has an infinite variance for samples within the mask and zero elsewhere.
The second row expresses the likelihood function as the marginalized conditional probability against an auxiliary variable $t$
\begin{eqnarray}
-2\ln\mathcal{P}(\tilde{d}, t | d) 
=&& t^T\mathbf{N}^{-1}_\mathrm{w}t 
+ (\tilde{d}-t)^T\mathbf{N}^{-1}_\mathrm{r}(\tilde{d}-t) \notag\\
&&+ (d-\tilde{d})^T\mathbf{N}^{-1}_\infty(d-\tilde{d}),
\end{eqnarray}
that has noise properties described by the white noise component of the noise model $\mathbf{N}_\mathrm{w}$ and its residual, $\tilde{d}-t$, that follows the red noise part $\mathbf{N}_\mathrm{r}=\mathbf{N}-\mathbf{N}_\mathrm{w}$ \citep{Huffenberger2018}. 
This formalism permits the ``regeneration'' of the gap-filled data by Gibbs sampling of the conditional probability functions. 
After 10 steps of sampling, we took the resultant $\tilde{d}$ as the updated version of the gap-filled data for another iteration of noise model estimation. 
Based on the Kullback–Leibler divergence between the result and simulation inputs with known noise properties, this iteration converged quickly most of the time when the mask fraction was low and when the masking was not correlated among the detectors. 
For some of the null tests \citepalias{eimer23} where the data were heavily masked (50\%) by splits between, e.g., positive and negative azimuth velocity scans, longer iterations were needed to obtain an accurate recovery of the noise model.
We chose to run this iteration five times for each noise model to accommodate the extreme cases.
The gap-filled data from Gibbs sampling were only used for updating the noise model and were not projected into the maps.

\subsection{Maximum-likelihood Mapmaking}\label{ssec:ML-map-making}
The maximum-likelihood solution to Equation \ref{eq:map-making-model} given the noise model $\mathbf{N}$ from Sections \ref{ssec:N-mod} and \ref{ssec:gap-filling} is
\begin{eqnarray}
    \hat{\mathbf{m}} = (\mathbf{P}^{T}\mathbf{N}^{-1}\mathbf{P})^{-1}\mathbf{P}^{T}\mathbf{N}^{-1}d.
    \label{eq:map-making}
\end{eqnarray}
This was solved iteratively using the preconditioned-conjugate gradient method \citep{painlessPCG}, where we used the inverse of the hits map as the preconditioner
\begin{equation}
    H^{-1} \equiv [(\mathbf{M}\mathbf{P})^T\mathbf{M}\mathbf{P}]^{-1}\label{eq:hits}. 
\end{equation}
Here, $\mathbf{M}$ and $\mathbf{P}$ are the modulation transfer function and the projection matrix defined in Equation \ref{eq:data-model}. 
This is a proxy for the covariance of the sky map assuming constant white noise in the raw data around the modulation frequencies. Using it as the preconditioner enabled fast convergence within 50 steps of conjugate gradient iteration.
Note that $\hat{\mathbf{m}}$ solved from Equation \ref{eq:map-making} is biased due to the filtering applied in Section \ref{ssec:map-filter}, and this bias will be characterized in Section \ref{ssec:map-transfer-function}.
The \texttt{healpix} pixelization at $N_\mathrm{side}=128$ was used for all map products at $40$~GHz, unless otherwise noted.

\begin{figure}
\includegraphics[width=\linewidth]{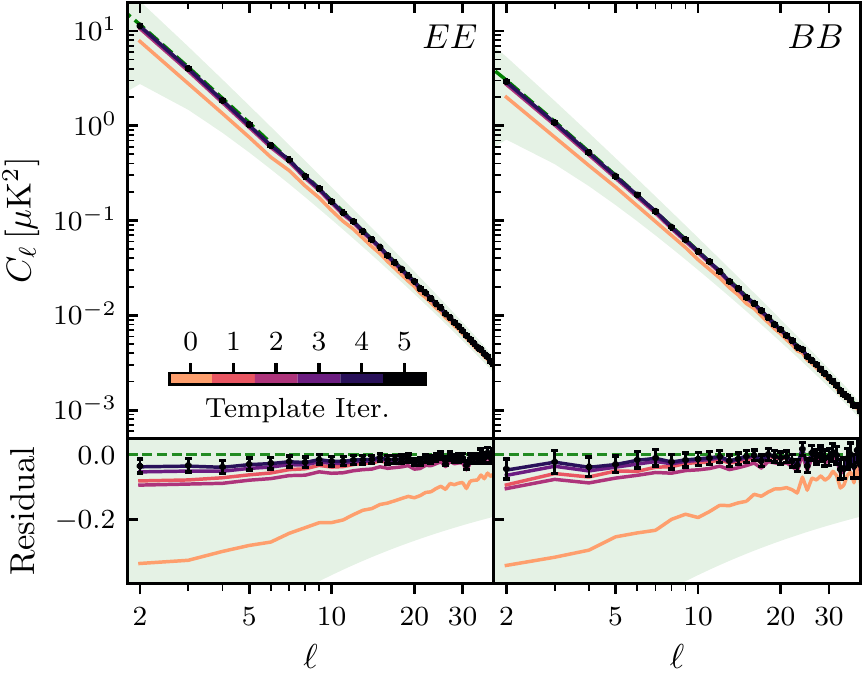}
\caption{The effect of template iteration in correcting the large angular scale power bias.
The top panels show the signal power spectra for $EE$ and $BB$. 
Green curves are the input synchrotron power spectrum model from \cite{planck18IV}, and the associated sample variance for a 75\% sky coverage. 
The colored curves show the estimated signal spectra, averaged over 2000 simulations, at each template iteration. 
The error bars are the standard error of the mean and are only shown for the last iteration for visual clarity.
The bottom panels show the residual spectra between the output spectra and the input spectra normalized by the input amplitude, in the same color scheme.
The correction converged quickly at high $\ell$ where the corresponding time streams are noise dominated; at low-$\ell$ the iteration converged by the fifth iteration, with a bias below $2.5\%$ for $\ell\leq5$, much smaller compared to the sample variance.}
\label{fig:template_iteration}
\end{figure}

The noise covariance estimate $\mathbf{N}$ in Section \ref{ssec:N-mod} has ignored the signal term in Equation \ref{eq:map-making-model}. 
While this is a good approximation, as the per-sweep data have very low S/N, 
it would slightly bias the signal map, 
especially on large angular scales, where the signal is degenerate with the correlated component in the noise model. 
To mitigate this bias, we performed multiple \textit{template iterations} by projecting out the estimated sky signal from the data for an updated noise estimation (as illustrated in Figure \ref{fig:pipeline-flowchart}, without additional Gibbs-sampling gap-filling). 
This technique has also been employed in experiments with similar mapmaking strategies \citep{dunner13, Aiola20, romero20}.
Figure \ref{fig:template_iteration} demonstrates this effect by showing the signal power spectra of the mapping pipeline, estimated from the cross-correlation of two independent splits at each template iteration.
The input signals were Gaussian realizations from power-law spectra following the best-fit synchrotron model of Planck \citep{planck18IV} with noise from the noise model estimated from the demodulated data. 
The bias on large angular scales was corrected for over a few template iterations; we found that using five iterations was sufficient for the 40~GHz maps with a remaining bias below $2.5\%$ for $\ell\leq5$. 
The remaining bias was verified to be caused by noise modeling since simulations using the input noise model (or other static noise models independent of the data) for weighting (i.e., Equation \ref{eq:map-making}) did not show this deficit of power at low $\ell$. 
The $40\,\mathrm{GHz}$ polarization maps did not suffer from the large-scale bias due to subpixel errors as pointed out in \cite{Naess22}, since CLASS demodulated data have a small dynamic range between large- and small-scale noise \citep[][\inprep{Cleary et al.}]{harrington21}; this prevents the subpixel residuals from outweighing the noise model at low frequencies and causing large-scale bias. 
This was verified by simulations that use higher-pixel-resolution sky maps as input. 

\subsection{Maps}\label{ssec:maps}
\begin{figure*}
\centering

\includegraphics[width=1\linewidth]{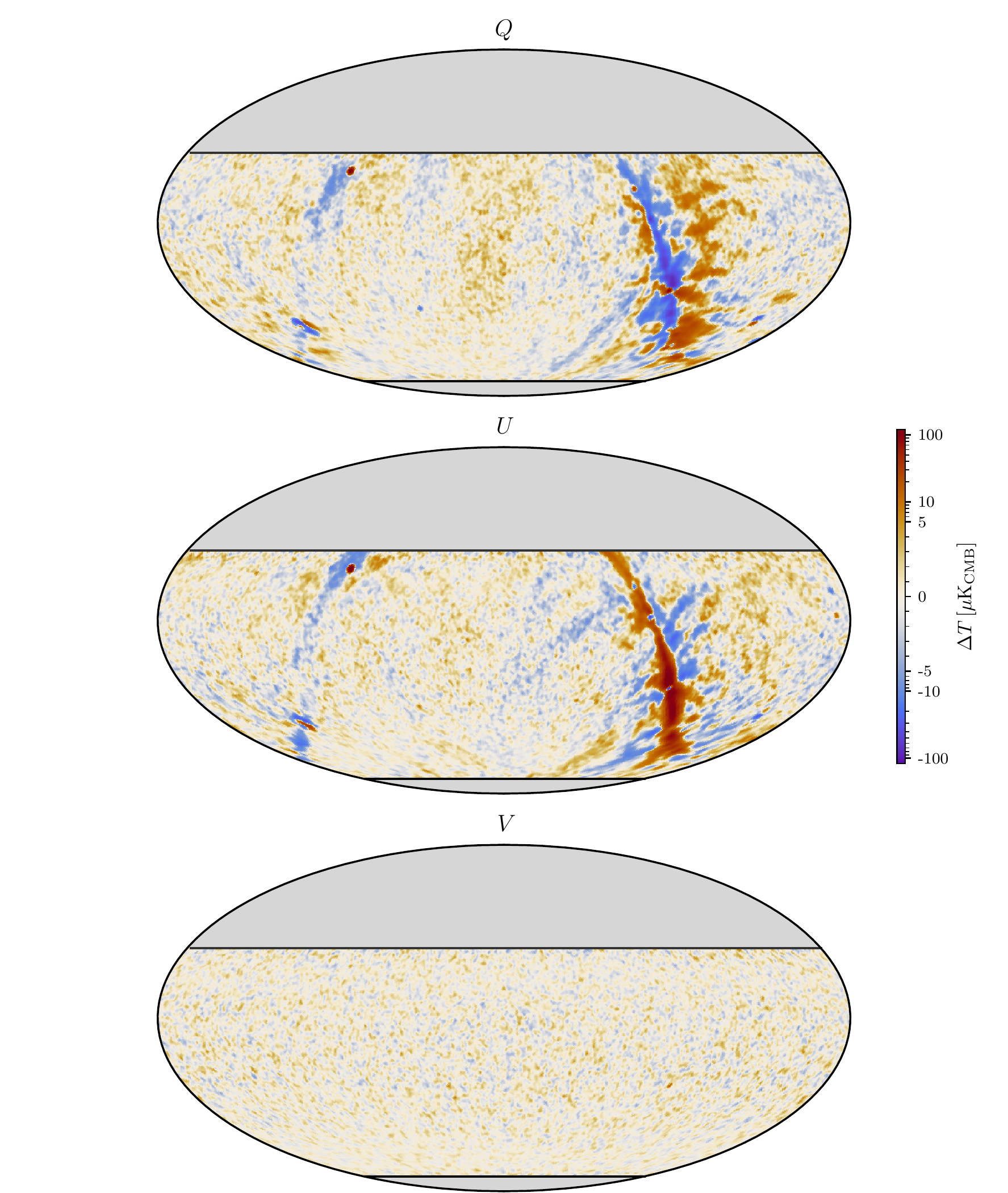}
\caption{
CLASS 40~GHz polarization maps in equatorial coordinates under Mollweide projection. 
The linear (Stokes $Q/U$) and circular polarization (Stokes $V$) are the final products of the data pipeline. The maps are smoothed to $2^\circ$ resolution to enhance the large-scale features. 
The gray-shaded regions are not surveyed. 
The color scale is linear below $5\,\mathrm{\mu K}$ and logarithmic above to show the structure in the map where bright synchrotron radiation dominates.
Due to the designed VPM throw, the noise levels are similar in the $Q$, $U$, and $V$ maps; therefore, the fluctuations in the $V$ map approximately represent the noise in the $Q/U$ maps. The apparent signals in the $Q/U$ maps are explored in \citetalias{eimer23}. }
\label{fig:maps}
\end{figure*}

\begin{figure*}
\includegraphics[width=1\linewidth]{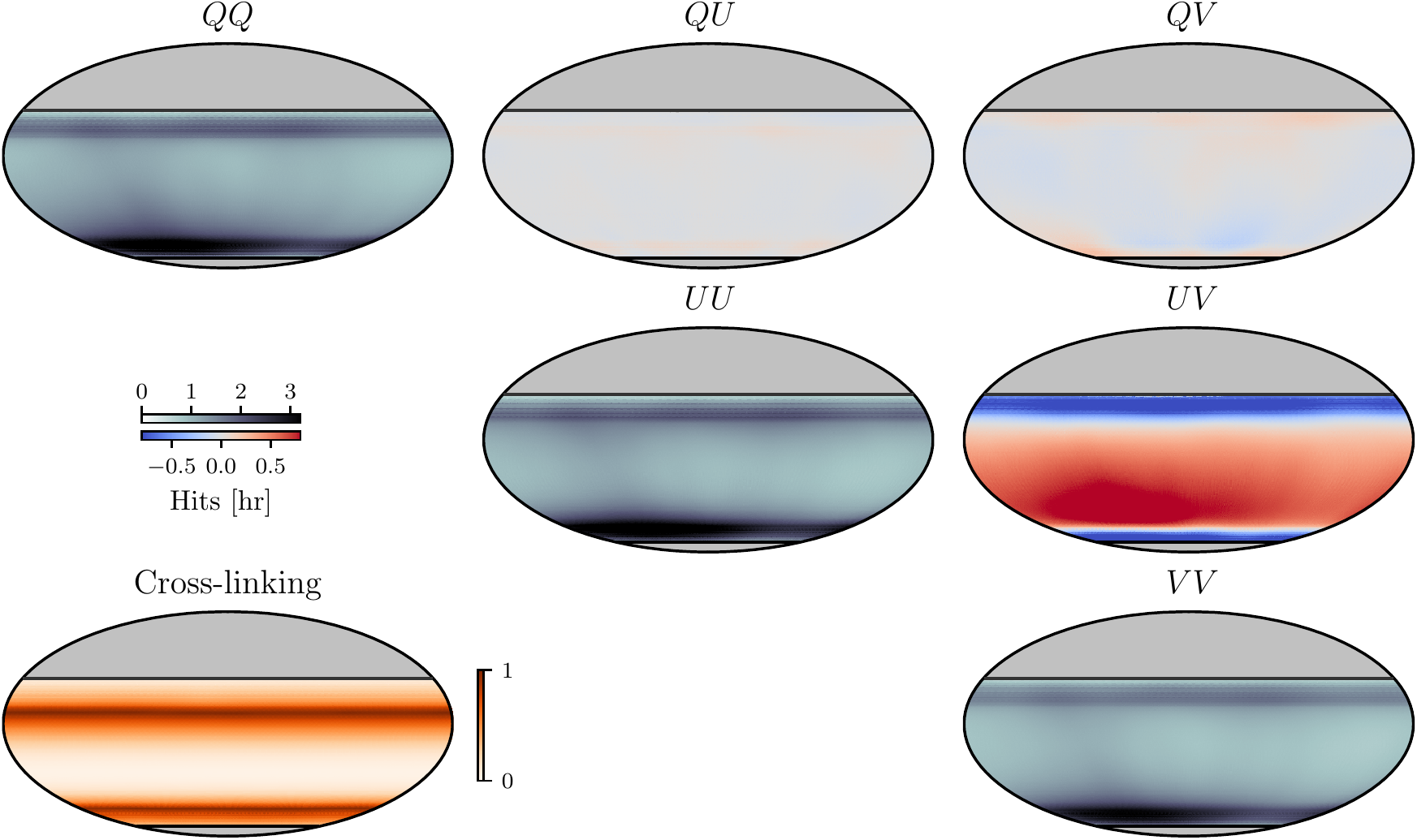}
\caption{CLASS 40~GHz hits maps and cross-linking map.
The upper triangle of the plot shows the hits maps in units of the integration time per $N_\mathrm{side}=128$ pixel ($0.46\,\mathrm{deg}^2$). This is proportional to the inverse-variance map assuming constant white noise in the raw data around the VPM modulation frequencies.
The bottom-left corner shows the cross-linking map in arbitrary units.
Higher values reflect even coverage of the scanning direction and therefore suppression of the scanning-related low-frequency noise.}
\label{fig:hits}
\end{figure*}

The 40~GHz linear and circular polarization maps made from the CLASS observations through 2022 are presented in Figure \ref{fig:maps}.
A battery of self-consistency null tests and a comparison with satellite missions will be presented in \citetalias{eimer23}.
In Figure \ref{fig:hits}, we show the hits count map defined in Equation \ref{eq:hits}.
The diagonal components are the integration time of each Stokes map; the off-diagonal terms reflect the covariance between maps. 
The $u$-$v$ covariance through the VPM modulation is integrated down in $QV$ but not in the $UV$ component due to the projection effect.\footnote{The covariance term between $Q$ and $V$ is symmetric between positive and negative boresight rotations and is thus canceled out in the total map, but the cancellation effect for $UV$ is smaller.}
The $QU$ component has minimal covariance due to the design of the CLASS scanning strategy.
In addition, the bottom-left corner of the figure is a cross-linking map from the CLASS scanning strategy, defined as 
\begin{equation}
1-\frac{1}{N}\left|\sum_{j}^{N}e^{-2i\Gamma_{j}}\right|,
\label{eq:xlinking}
\end{equation}
where the sum is over the $N$ TOD falling within a pixel, and $\Gamma_j$ is the angle between the scanning direction with respect to the local meridian for the $j\,\mathrm{th}$ sample \citep{Aiola20, mccallum21}.
This value reflects the uniformity of the scanning direction coverage and is a good proxy for the (inverse) large-scale noise related to atmospheric emissions \citep{atkins23}.
Together, the cross-linking and hits maps show complementary information about the CLASS sensitivity on the sky at different scales.

The noise power spectra estimated from simulations are shown in Figure \ref{fig:systematics-Cl}, which reach white noise levels 
$110\,\mathrm{\mu K\,arcmin}$ in $EE/BB/VV$.
After correcting for the mapping transfer function, the large-scale noise has a logarithmic slope close to $-2.4$ and knee angular scales (at which the spatially correlated noise equals the white noise) of $\ell=12$ and $18$ for circular and linear polarization, respectively.

\subsection{Mapping Transfer Function and Reobservation}\label{ssec:map-transfer-function}
\begin{figure*}
\includegraphics[width=\textwidth]{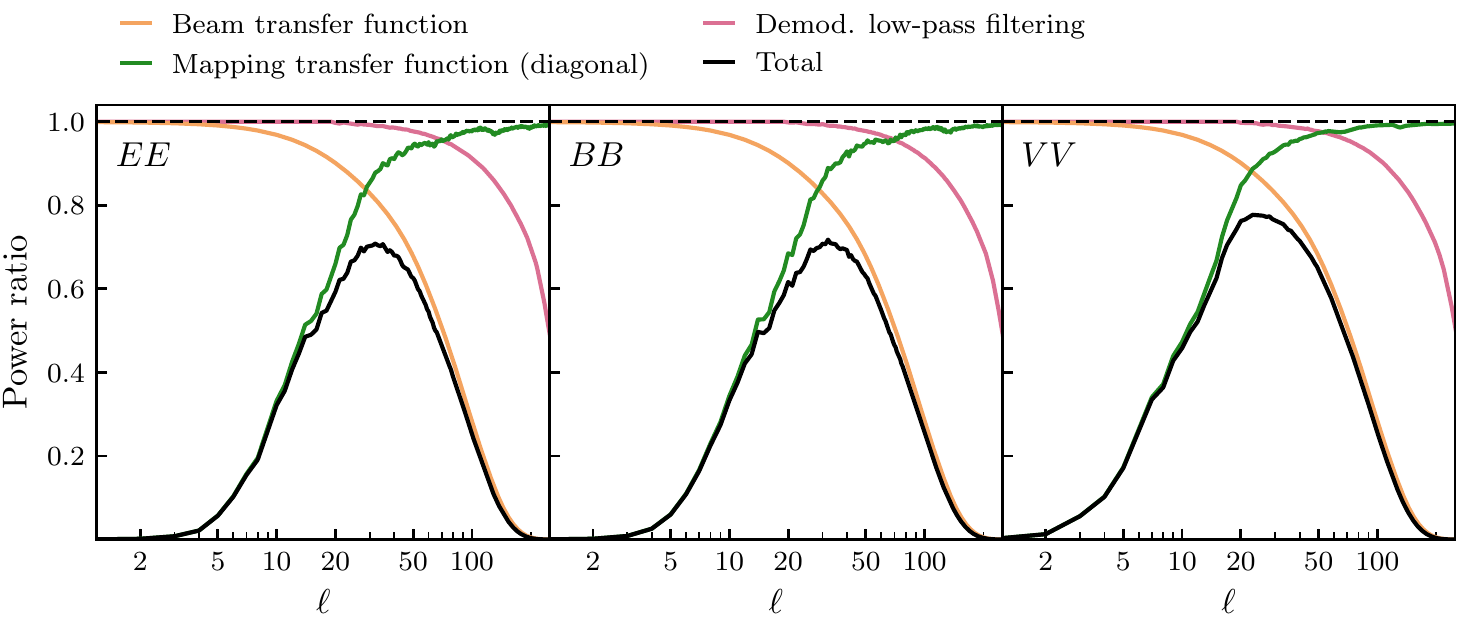}
\caption{
The transfer functions for linear polarization $EE/BB$ and circular polarization $VV$. 
The effects of the demodulation low-pass filter
are shown in pink{\marker{.82}{.35}{.51}}, and the diagonal components of the mapping transfer matrix $F_{\ell\ell}$ are in green{\marker{.12}{.49}{.10}}. Together with the beam window function \citep[][orange curve{\marker{.95}{.62}{.37}}]{xu20}, they make up the total transfer function of the signal (black{\marker{0}{0}{0})}.}
\label{fig:filtering_tf}
\end{figure*}

\subsubsection{Harmonic-domain Transfer Matrix}
The filtering performed on both raw data (Section \ref{ssec:demodulation}) and demodulated data (Section \ref{ssec:map-filter}) removed power from the sky signal.
This effect can be modeled as
\begin{equation}
    C_\mathrm{\ell,\mathrm{out}} = F_{\ell\ell'} C_{\ell', \mathrm{in}},
\end{equation}
where $\ell$ and $\ell'$ are extended multipole indices that run through multipole moments of $\{VV, EE, BB\}$, and $F_{\ell\ell'}$ is the \textit{mapping transfer matrix}.
Here we have assumed isotropic filtering on each mode, which is only a good approximation for statistically isotropic sources at small angular scales. Despite this, we found through simulations that the resulting harmonic transfer function results in unbiased spectra for $\ell>4$.
Estimation of the filter transfer matrix was obtained by mapping signal simulations with known input spectra, performing the same filtering and noise weighting as the data, and comparing the resultant spectra.
The off-diagonal components, i.e., mode mixing among adjacent multipoles and among Stokes parameters, were found to be mostly insignificant; only the covariance over $10$ adjacent multipoles in the $EE$ and $BB$ blocks and five adjacent multipoles below $\ell=30$ in the $EE-BB$ cross blocks were modeled, and the remaining elements were fixed to zero.

Figure \ref{fig:filtering_tf} shows the transfer function due to low-pass filtering in demodulation (Equation \ref{eq:demod_solution}) and the diagonal component of the mapping transfer matrix $F_{\ell\ell}$. For linear polarization, about $35\%$ of the power is retained at $\ell=10$, and the signal sensitivity peaks at $\ell\approx40$ (after accounting for the beam window function and noise power spectrum).

\subsubsection{Pixel-space Transfer Matrix}
At large angular scales, the anisotropic effect of both the foreground signals and filtering is more prominent, and the harmonic-domain transfer matrix is a less-robust representation. 
A pixel-space transfer matrix can be introduced at low resolution for this situation:
\begin{equation}
    m_{i,\mathrm{out}} = F_{ij} m_{j, \mathrm{in}},
\end{equation}
where $m_{\mathrm{in/out}}$ are the input and filtered maps downgraded to $N_\mathrm{side}=16$ resolution, $F_{ij}$ is the transfer matrix estimated from an ensemble of signal-only simulations, and the subscripts denote the map pixel index.
Although the mapmaking pipeline is linear, this equation is only an approximation for the downgraded maps due to the noncommutativity of the mapping and the downgrade operation and showed a $\sim10\%$ discrepancy at the largest angular scales when compared to the reobservation (Section \ref{ssec:reobs}). 
This can be improved in the future by using a separate low-resolution mapping pipeline.

This transfer matrix can be used for pixel-space analyses, and it can be integrated with quadratic estimators \citep[e.g.,][]{xQML} to optimally correct for the bias in the power spectra.
For the latter, we found that the transfer-function-corrected power spectra are unbiased for $\ell\geq4$, comparable to the pseudo-$C_\ell$ estimator with the harmonic-domain transfer function correction, but that they are statistically more optimal at low-$\ell$ than the pseudo-$C_\ell$ approach.

\subsection{Reobservation}\label{ssec:reobs}
To facilitate direct comparison with other experiments, in particular the all-sky maps from WMAP and Planck, we applied the CLASS filtering and weighting on these maps to forward-model the filtering effects.
The reobservation started with convolving the input map to $1.5 ^\circ$ (FWHM) resolution.
The linear polarization components of the input map were then projected to the CLASS \emph{demodulated time streams}\footnote{In principle, the reobservation should go through the modulation and demodulation process as well; however, in practice, it is much more efficient to start the simulation from the demodulated stage. Although the filtering effect due to the low-pass filtering in the demodulation (the pink curve in Figure \ref{fig:filtering_tf}) is not captured, its effect is subdominant compared to the beam transfer function and is safe to be neglected.} using the CLASS pointing model, and the circular polarization time streams were set to zero.
These data were then filtered in the same ways as the CLASS demodulated data and projected back to the maps using the fixed noise model from the last template iteration of the CLASS mapmaking procedure---no template iteration was performed for reobservation.
Since all mapping operations are linear, this is an accurate description of the filtering that the CLASS data have undergone.

\section{Data validation and Systematics}\label{sec:systematics}
\subsection{Systematics and Simulation}
\begin{figure*}
    \includegraphics[width=\linewidth]{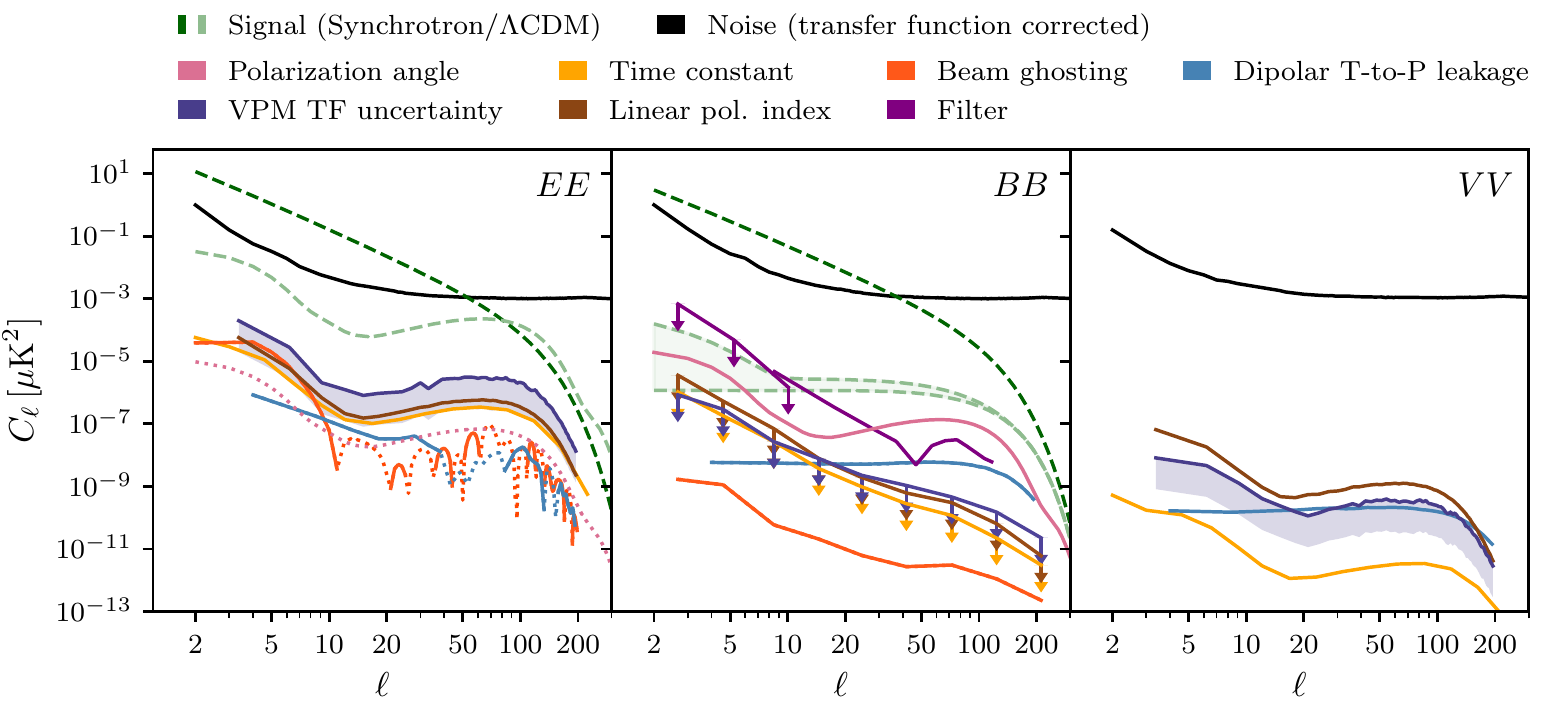}
    \caption{
        \label{fig:systematics-Cl}
        Summary of the effect of multiple systematic errors on the power spectra.
        The black curve{\marker{0}{0}{0}} shows the map noise estimated from an ensemble of simulations with the mapping transfer matrix corrected.
        For reference, the signal spectra are plotted as 
        dashed curves for the diffuse synchrotron signal at 40~GHz \citep[][dark green{\marker{0.04}{0.33}{0}}]{planck18IV} and the CMB \citep[][light green{\marker{0.49}{0.69}{0.49}}, a range of tensor-to-scalar ratio $0<r<0.01$ is represented by the shaded green area in the $BB$ panel]{planck18VI}, respectively. 
        The signal spectra are convolved with the $1.5^\circ$ FWHM beam window function.
        The solid/dotted curves are the measured positive/negative systematic bias; the curves with downward arrows represent the $2\sigma$ confidence level upper limit for systematics that are not detected given the sample variance in the simulations.
        \textit{Pink}{\marker{0.82}{0.35}{0.51}}: systematic error from the polarization angle calibration uncertainty (Section \ref{ssec:sys-pol-angle}).
        \textit{Orange}{\marker{0.99}{0.58}{0.04}}: effect of a $0.1 \,\mathrm{ms}$ bias in the detector time constants (Section \ref{ssec:sys-tau}).
        \textit{Red}{\marker{0.99}{0.25}{0.09}}: effect of a $3\times10^{-3}$ level beam ghosting across the focal plane (Section \ref{ssec:sys-ghost}).
        \textit{Light blue}{\marker{0.22}{0.43}{0.65}}: effect of the dipolar $T$-to-$P$ leakage (Section \ref{ssec:sys-dipole-t2p}).
        \textit{Navy blue}{\marker{0.22}{0.16}{0.47}}: uncertainty in the VPM transfer function parameters. The curves indicate the maximum variation in the residual power spectra for VPM parameters drawn from the $2\sigma$ confidence interval of the VPM parameter optimization process (Section \ref{ssec:sys-vpm-tf}). 
        The shaded regions in $EE$ and $VV$ highlight that these are the result of an ensemble of parameters, but are not quantitative depictions of the spread.
        \textit{Brown}{\marker{0.47}{0.20}{0.07}}: effect of a $+0.3$ bias in the linear polarization spectral index assumed for demodulation (Section \ref{ssec:sys-vpm-tf}). 
        \textit{Purple}{\marker{0.42}{0.}{0.43}}: residual $E$-to-$B$ mixing due to the mapping filters after the transfer matrix correction (Section \ref{ssec:sys-filter}). 
    }
\end{figure*}

In this section, we characterize several types of systematic errors and assess their impact on the scientific result.
These issues were studied through simulations with CMB realizations as input, and the resultant bias to the power spectrum was characterized by the difference between the systematics-included auto-power spectrum and the input power spectrum. 
The simulations were drawn from the Planck best-fit parameters \citep{planck18I} with the $B$-mode amplitudes set to zero.
Since the input had no power in the $B$ mode and in circular polarization, the effects in $BB$ and $VV$ were dominated by the auto-correlation of the systematics residuals, i.e., a second-order effect of the systematics, while the residual $EE$ spectrum were dominated by the cross-correlation between the residual $E$ model power and the original signal, i.e., a first-order effect.
Figure \ref{fig:systematics-Cl} summarizes the results, which we describe below.

\subsubsection{Detector Polarization Angles}\label{ssec:sys-pol-angle}
Calibration of the absolute polarization angle is critical for accurate separation of the $E/B$-mode signal \citep{hu03} and the search for parity-violating physics \citep{Finelli09}.
Systematic uncertainties associated with the alignment of individual detector pixels,  offsets between the focal plane and the VPM wire, pointing errors in the telescope boresight rotation, and modeling errors in the optics
can lead to bias in the polarization angle.

We used the bright polarized source Tau A as the main calibrator whose polarization angle was measured by CLASS to be $-87.02\pm0.2^\circ$ in Galactic coordinates\footnote{The polarization angles in this paper follow the IAU convention.} where the statistical error is derived from noise simulations.
However, as shown in Figure \ref{fig:wire-dir}, only the central bottom part of the focal plane covered Tau A at all boresight rotations; detectors on the sides observed Tau A only at certain boresight rotations, and part of the top detectors never saw Tau A.
This boresight-dependent partial coverage of Tau A limits its ability to characterize systematic errors of the polarization angle.
Based on the optics of the telescope and the distribution of the wire direction $\phi_P$ (Equation \ref{eq:wire-angle}) across the focal plane, we used the discrepancy in the Tau-A polarization angle measured between splits of the data to assess the systematic errors.
The ``quadrupole'' split defined by the sign of $\phi_P$ \citepalias[see details in][]{eimer23}
probes the systematic effects in the optics modeling.
Similarly, a split between scans with positive and negative boresight rotation relies on either side of the blue detectors in Figure \ref{fig:wire-dir} to measure Tau A and is therefore sensitive to the optics model as well.
The Tau A polarization angle differences in these two ways of splitting the data are $0.70^\circ$ (quadrupole) and $0.80^\circ$ (boresight).
The Tau A angle measured from the data using each of the three VPM grids shows a maximum discrepancy of $0.37^\circ$, which indicates the level of the error caused by an angular offset between the VPM grid and the focal plane.
Combining these two factors, we assign $0.7^\circ$ to the systematic error in the polarization angle calibration.
The final measurement of Tau A polarization angle from CLASS, $-87.02\pm0.20\,(\mathrm{stats.})\pm0.70\,(\mathrm{sys.})^\circ$, is consistent with that from WMAP \citep[$-87.3\pm0.2\pm1.5^\circ$,][]{jarosik07,weiland11} and Planck \citep[$-88.65\pm0.79\pm0.50^\circ$,][]{planck15XXVI} at similar frequencies.
At around $40\,\mathrm{GHz}$, the CLASS beam at FWHM $1.54^\circ$ is wider than that of WMAP ($0.49^\circ$) and Planck ($0.47^\circ$), and the resolution confusion can contribute a subdegree discrepancy in the polarization angle determination.
Since no significant discrepancy in the polarization angle is found within internal comparison or externally with other experiments \citep{aumont20}, we do not apply any correction to the detector angle.

The $E$-$B$ mixing from a $0.7^\circ$ polarization angle error is shown in Figure \ref{fig:systematics-Cl} as the pink curve based on the analytical model \citep{keating13}.

\subsubsection{Time Constants}\label{ssec:sys-tau}
The time constants used for data reduction are the medians of values estimated from each DataPkg per detector per observation era. So for each detector, there is a single time constant estimate used for all data in Era~1 and another for Era~2.
The time constants among the detectors have a typical value of $3\,\mathrm{ms}$ (median), but are sensitive to the optical loading from the sky and show correlations with the air temperature, PWV, and the telescope boresight rotation.
Most notably, air temperature accounts for a shift of $0.2\,\mathrm{ms}$ (comparable to the standard error of the time constant estimations) in the time constants of all detectors over its normal range $(-10\mathrm{^\circ C},\,6\mathrm{^\circ C})$.
The time constants are also affected by the thermal history of the detectors; some detectors jump between states of time constants differing by approximately $0.2\,\mathrm{ms}$ when the focal plane temperature warms up above $0.1\,\mathrm{K}$.
In the following analysis, we take half the value of $0.1\,\mathrm{ms}$ to study the impact of a systematic bias. The statistical uncertainties from averaging all time constants per era are insignificant compared to this.

These variations are not considered in the pipeline, and 
deconvolution with biased detector time constants has dual impacts on the CLASS data. 
First, the actual pointings of the telescope would be offset from the calibrated encoder values, but this effect is negligible compared to the beam scale of the telescope and is further diminished by the forward and backward scanning of the telescope.
More importantly, the phase delay from the VPM encoder due to the biased time constants would cause leakage between the linear and circular polarization. 
Figure \ref{fig:systematics-Cl} shows the effect of the time constants biased by $0.1\,\mathrm{ms}$. 
This is an approximately $10^{-3}$ effect in the mixing of the polarization states; therefore, for simulation with pure $E$ mode input, the residuals in the $EE/VV$ power spectra are at the $10^{-3}$/$10^{-6}$ level, respectively.
No $E$-to-$B$ leakage is detected above the level of the lensing $B$-mode at the largest angular scales.

\subsubsection{Ghosting Beam}\label{ssec:sys-ghost}
Beam ghosting caused by the internal reflection of the telescope is detected in Moon observations at a level $3\times10^{-3}$ of the main beam at the opposite position of the focal plane for each detector. 
To simulate this effect, a Gaussian beam centered on the opposite point of the focal plane was assigned to each detector, with the peak amplitude of the beam consistent with the reflection amplitude measured from the Moon maps. 
These ghosting beams were then convolved with the sky simulations for mapping.
The residual power spectra of the ghosting beam, which are shown in red in Figure \ref{fig:systematics-Cl}, have the greatest impact at angular scales greater than the field of view of the telescope ($\gtrsim20^\circ$).

\subsubsection{Temperature-to-polarization Leakage}\label{ssec:sys-dipole-t2p}
\begin{figure}
\includegraphics[width=\linewidth]{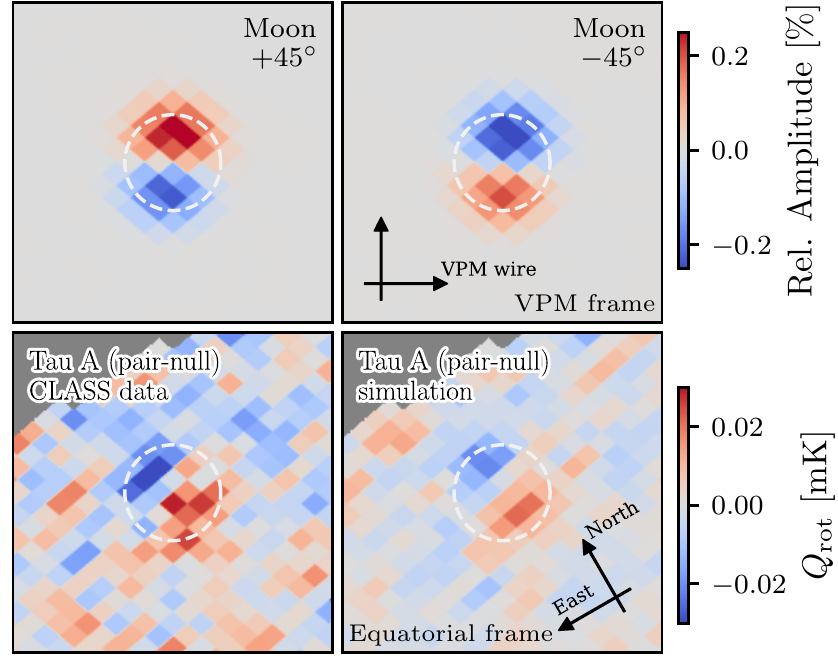}
\caption{
        Effect of dipolar $T$-to-$P$ leakage. 
        \textit{Top}: the demodulated linear polarization (Stokes $u$) Moon maps in the VPM coordinates for a pair of detectors. In this coordinate system, the VPM wire grid is horizontal. The colors are scaled to the peak amplitude of the Moon temperature maps.
        The $1.54^\circ$ FWHM main beam is marked by the white dashed circle.
        \textit{Bottom left}: the differential linear polarization maps of Tau A between all $+45^\circ$/$-45^\circ$ detectors. 
        The maps are rotated by the Tau A polarization angle at 40~GHz \citep{weiland11} so that any residual Tau-A polarization signal would appear entirely as a point source in the Stokes-$Q$ map (denoted $Q_{\mathrm{rot}}$). Because the $+45^\circ$/$-45^\circ$ signals are differenced, their opposite-signed dipoles (top row) average constructively. 
        \textit{Bottom right}: simulation of the dipolar leakage effect from convolving scaled WMAP $Q$-band temperature maps with the dipolar leakage beam from all detectors.
    }
    \label{fig:dipole-beam}
\end{figure}

The placement of VPM as the first element in the optical path is to prevent polarization due to oblique reflection from being modulated. 
Moon maps made from dedicated Moon scans \citep{xu20} have shown that the monopole $T$-to-$P$ leakage is at the $4\times10^{-5}$ level for pairs of detectors. 
The Moon maps also reveal a dipolar pattern that takes opposite signs in linear polarization for a pair of detectors with its amplitude and orientation independent of the telescope boresight rotation (top row of Figure \ref{fig:dipole-beam}).
Similar patterns are also observed for circular polarization, but at lower magnitudes and with the orientations of the dipole offset by approximately $45^\circ$ from those in linear polarization.
This effect is consistent with a misalignment of the VPM, where the tilt between the grid and the mirror creates differential pointing and leads to an additional term in Equation \ref{eq:mono-modulation-function} proportional to the brightness temperature gradient along the tilt direction. 
This term is modulated at the VPM frequency and is covariant with the linear (primarily) and circular polarization modulation function and is picked up by demodulation \citep[Section 4.1.2 of][]{harrington2018thesis}. 

The bottom-left panel of Figure \ref{fig:dipole-beam} shows the pair-null linear polarization map made by differencing polarization maps made with the $+45^\circ$ detectors from those made with the $-45^\circ$ detectors. Because of polarization modulation, the linear polarization can be recovered with the $+45^\circ$ and $-45^\circ$ detectors separately. Thus, differencing removes the polarization signal, and enhances the oppositely signed dipolar $T$-to-$P$ leakage (Figure \ref{fig:dipole-beam}, top row). Furthermore, the $T$-to-$P$ effect can be simulated by convolving the sky temperature signal with the dipolar beam estimated from the Moon maps for each detector and each VPM grid.  
The convolution was performed in the pixel space using \texttt{pisco} \citep{fluxa20}. The bottom-right of Figure \ref{fig:dipole-beam} shows a simulation of the bottom-left panel made by convolving the WMAP $Q$-band temperature maps scaled to the CLASS bandpass by the dipolar leakage beams and differencing the simulated $+45^\circ$ and $-45^\circ$ leakage maps. The agreement between the data and simulation shows that the dipolar leakage measured from the moon is in agreement with that measured in the CLASS survey maps via Tau A.

Although the dipolar leakage is on average $0.3\%$ compared to the main beam for a single detector, its impact on the final maps is further diminished when polarization data from pairs of $+45^\circ$/$-45^\circ$ detectors are averaged, and the dipoles in the top row of Figure \ref{fig:dipole-beam} cancel each other (instead of reinforce, as in the bottom row). Only eight of the 72 detectors in Era 1 were unpaired (due to readout failures), six (two) of which are $+45^\circ\,(-45^\circ)$ oriented, and this was reduced to a single unpaired detector in Era 2 (due to data selection).

To assess the impact of this leakage in the angular power spectrum, we convolved CMB temperature map simulations with the dipole beam in linear and circular polarization. 
We then took cross spectra between the output maps and the sum of the input and output (i.e., the main beam plus the dipole beam). 
The resultant $EE$ power spectrum has a contribution from the auto-correlation of the dipole systematics and the cross-correlation with the CMB $E$ mode due to the CMB $TE$ correlation; the effects on the $BB/VV$ power spectra are solely from the auto-correlation.
These results are shown in Figure \ref{fig:systematics-Cl} as blue curves, and in all cases, the effects of the dipole $T$-to-$P$ leakage are subdominant compared to the noise level and/or the cosmology signals of interest.

\subsubsection{VPM Transfer Function Uncertainty}\label{ssec:sys-vpm-tf}
The best-fit VPM transfer function parameters were determined with a combination of instrument characterization and a polarization-leakage minimization process as outlined in Section \ref{ssec:vpm-parameters}.
We assessed the impact of VPM parameter uncertainties by modulating a single realization of the sky signal with different VPM parameters drawn from the likelihood chain in Section \ref{ssec:vpm-parameters} that are within the $95\%$ confidence region around the best fit.
These simulations were then demodulated with the best-fit VPM parameters and mapped in the same way as the data.
The maximum absolute differences between the output and input power spectra are plotted in Figure \ref{fig:systematics-Cl} as the navy blue curves with shades. 
The VPM parameter uncertainties typically translate to a $1\%$ error in map amplitudes.

The demodulation pipeline assumed a single spectral index for the linear polarization, which is a simplification to the real-world case where both the spectral index of the dominant synchrotron emission and the mixing between different components vary across the sky. The \texttt{PySM} \citep{pysm} simulation with realistic input from synchrotron, CMB, and dust suggests that the aggregated effect corresponds to a standard deviation across the sky of $0.3$ at $40\,\mathrm{GHz}$.
Figure \ref{fig:systematics-Cl} shows in brown a simulation with a uniform $+0.3$ bias in the linear polarization spectral index; the leakage effect manifests as a transfer of the $E$ mode power into circular polarization.
This result should be considered conservative since the variation of the index bias over the sky should partially cancel out and leave less residual in the power spectra.

\subsubsection{Filtering Artifacts}\label{ssec:sys-filter}
The demodulated data filtering performed before mapmaking removes and redistributes the sky signal over large angular scales. 
Although its effect in the harmonic domain has been modeled by the transfer matrix (Section \ref{ssec:map-transfer-function}), the insufficiency in the modeling could still lead to bias in the corrected power spectra. 
The purple curve in the $BB$ panel of Figure \ref{fig:systematics-Cl} shows the transfer function-corrected $BB$ power spectra from filtered $E$-mode-only simulations. 

\subsection{Internal Consistency Test}\label{ssec:null-result}
The validation of the CLASS data product is checked through a series of internal consistency tests. 
The tests are executed by splitting the demodulated data into two similar-sized subsets (denoted A and B) that are expected to expose certain types of systematic error. Two homogeneous temporal split maps are made for each A/B split through the same mapmaking pipeline.
The cross spectra of the difference (null) maps $\mathrm{A}-\mathrm{B}$ from the two temporal splits are computed and compared to an ensemble of simulations to check for consistency.
Details of the design of the split and the systematic errors probed by each null test, as well as the final results, are presented in \citetalias{eimer23}.

\section{Conclusion}\label{sec:conclusion}
We have presented a detailed description of the CLASS data reduction pipeline for 40~GHz observations conducted from August 2016 to May 2022. When weather, instrument upgrades, and other interruptions permitted, observations were conducted continuously, regardless of time-of-day or season-of-year. After all data cuts, the analysis incorporated 86.77 detector-years of data, representing $\sim20$\% of the possible data volume. These data cover 75\% of the sky and extend from $-76^\circ$ to $30^\circ$ in declination. 

The sky polarization signal in the data was amplitude-modulated at the VPM frequency ($10\,\mathrm{Hz}$) and its harmonics. Therefore, the selected data were filtered and demodulated to remove the dominant time-correlated noise in the raw data below $5\,\mathrm{Hz}$ and retain the polarization signal in the $0.5-1\,\mathrm{Hz}$ wide side-bands around the modulation harmonics. Isolating the polarization signal from the correlated noise in this way is the most important aspect of the CLASS strategy for achieving the large-angular-scale measurement. The demodulated polarization data were then filtered to remove systematic effects, such as azimuth-synchronous and wind-induced signals. The type and level of filtering were tuned to enable the data to pass internal consistency ``null'' tests \citepalias{eimer23}. The demodulated and filtered data were then input to an iterative preconditioned-conjugate-gradient-descent algorithm to jointly solve for the maximum-likelihood Stokes $Q$, $U$, and $V$ maps. Due to the filtering, the maps are biased low on large angular scales. We used simulations to show that the bias to the angular power spectra is $\sim67\%\,(\sim85\%)$ at $\ell=20$ and $\sim35\%\,(\sim47\%)$ at $\ell=10$ for linear (circular) polarization. After correcting this bias, the noise level in the angular power spectra was found in data-based simulations to be $110\,\mathrm{\mu K\,arcmin}\,[1+(\ell_{\mathrm{knee}}/\ell)^{2.4}]^{1/2}$, with $\ell_{\mathrm{knee}}\approx12$ for circular polarization and $\ell_{\mathrm{knee}}\approx18$ for linear polarization. With these maps, CLASS is pushing the limits of what has been achieved from a suborbital platform at the largest angular scales.

Multiple sources of systematic error were quantitatively studied with simulations. The bias induced in the $\Lambda\mathrm{CDM}$ $EE$ angular power spectrum was found to be subpercent. Leakage from the $EE$ to $BB$ spectra was found to be comparable to the predicted $B$ mode spectrum with $r=0.01$. Improvements in calibration and the data pipeline will reduce the leakage. For CLASS, the $40\,\mathrm{GHz}$ data are intended to measure the synchrotron foreground and not to constrain $\Lambda\mathrm{CDM}$. Therefore, this study of the impact of systematic errors on the $\Lambda\mathrm{CDM}$ spectra is provided as an initial benchmark of the analysis on the way to analyzing the multifrequency dataset. Additionally, the subpercent bias found for the $\Lambda\mathrm{CDM}$ $EE$ spectrum should be similar to the expected bias level for the $EE$ and $BB$  spectra of the diffuse synchrotron emission.

This is the first demonstration of the full data pipeline for CLASS.
At the time of writing, the methods developed here for demodulation and mapping were being applied, adapted, and improved on data from the other CLASS frequency bands. Several hardware improvements were made (guided by data), and both software and further hardware improvements were desired and planned. These results are therefore preliminary. Together with previous, ongoing, and planned improvements to the instrument and measurement strategy, future analyses will provide an independent view of the CMB polarization at the largest angular scales.

\section{Acknowledgments} 
We acknowledge the National Science Foundation Division of Astronomical Sciences for their support of CLASS under grant Nos. 0959349, 1429236, 1636634, 1654494, 2034400, and 2109311.
We thank Johns Hopkins University President R. Daniels and Dean C. Celenza for their steadfast support of CLASS. We further acknowledge the very generous support of Jim and Heather Murren (JHU A\&S '88), Matthew Polk (JHU A\&S Physics BS '71), David Nicholson, and Michael Bloomberg (JHU Engineering '64). The CLASS project employs detector technology developed in collaboration between JHU and Goddard Space Flight Center under several previous and ongoing NASA grants. Detector development work at JHU was funded by NASA cooperative agreement 80NSSC19M0005. 

We acknowledge scientific and engineering contributions from Max Abitbol, Fletcher Boone, David Carcamo, 
Manwei Chan, 
Joey Golec, Dominik Gothe, 
Ted Grunberg, 
Mark Halpern, 
Saianeesh Haridas, 
Kyle Helson, Gene Hilton, 
Connor Henley, 
Johannes Hubmayr, 
Lindsay Lowry, 
Jeffrey~John McMahon, 
Nick Mehrle, 
Carolina~Morales Perez, 
Ivan~L. Padilla, Gonzalo Palma, Lucas Parker, 
Bastian Pradenas, Isu Ravi, 
Carl~D. Reintsema, 
Gary Rhoades, Daniel Swartz, Bingjie Wang, Qinan Wang, Tiffany Wei, and Zi\'ang Yan. We thank Miguel Angel D\'iaz, Joseph Zolenas, Jill Hanson, William Deysher, Mar\'ia Jos\'e Amaral, and Chantal Boisvert for logistical support. We acknowledge the productive collaboration of the JHU Physical Sciences Machine Shop team. 
S.D. is supported by an appointment to the NASA Postdoctoral Program at the NASA Goddard Space Flight Center, administered by Oak Ridge Associated Universities under contract with NASA.
R.D. thanks ANID for grants BASAL CATA FB210003 and FONDEF ID21I10236.
R.R. is supported by ANID BASAL grant FB210003.

Part of this research project was conducted using computational resources of Advanced Research Computing at Hopkins (ARCH) and the National Energy Research Scientific Computing Center (NERSC). CLASS is located in the Parque Astron\'omico Atacama in northern Chile under the auspices of the Agencia Nacional de Investigaci\'on y Desarrollo (ANID).

\software{
numpy \citep{numpy20}, 
scipy \citep{scipy}, 
scikit-learn \citep{scikit-learn},
matplotlib \citep{matplotlib},
astropy \citep{astropy}, 
healpix \citep{healpix},
getdata \citep{getdata},
pyephem \citep{pyephem}, 
camb \citep{camb},
pysm \citep{pysm},
polspice \citep{polspice},
emcee \citep{emcee},
}

\bibliographystyle{aasjournal}
\bibliography{cosmology,references, software_common, class_common, cmb, foreground, hardware}
\appendix
\section{Pointing Model}\label{sec:pointing-model}
The CLASS pointing model is a 34-parameter model for determining encoder positions to achieve a desired pointing in azimuth, elevation, and boresight angle. We present the model here, as it is unique in how it handles errors inherent to a three-axis mount. For a mount with one telescope, there is one pointing model, and the mount is positioned such that the commanded position corresponds to array center in azimuth and elevation at a boresight angle with respect to the zero boresight angle of the array. For a mount with two telescopes, each telescope has its own pointing model. While the individual models are used in the data analysis, the mount uses the average of the two for positioning such that the commanded position corresponds to a point on the sky halfway between the two array centers at a boresight angle with respect to a zero angle halfway between the zero boresight angle of the two arrays. Henceforth, when referring to the telescope mount, this average position will be denoted as the array center.

\subsection{Boresight Pointing Model}

The telescope mount boresight is defined as the axis of rotation of the boresight platform that houses the telescopes. As the boresight azimuth, elevation, and rotation angle, as read from the encoders, are not perfectly aligned with the sky, we need a pointing model to correct this misalignment. We use a boresight pointing model that contains 21 terms: 11 in azimuth and 10 in elevation shown in Table \ref{tab:boresight}. Each term corresponds to a physical effect that affects the alignment of the boresight. Of these 21 terms, 12 are currently in use; seven in azimuth and five in elevation. The four tilt terms are reduced to two coefficients in the pointing model data reduction, as they are not independent and are used in linear combination to describe the tilt of the mount as a rotation:

\begin{align}\label{eq:mount-tilt}
\Delta \mspace{2mu} az &= \alpha\sin(el)\cos(az) + \beta\sin(el)\sin(az) \\
\Delta \mspace{2mu} el &= \beta\cos(az) - \alpha\sin(az),
\end{align}

\noindent where $\alpha$ represents a tilt of the mount to the West and $\beta$ represents a tilt to the North. Here $az$ and $el$ are the commanded position of the mount boresight and $\Delta \mspace{2mu} az$ and $\Delta \mspace{2mu} el$ are the tilt-related pointing corrections in azimuth and elevation. All azimuth pointing corrections are in units of true arc on the sky and are subsequently multiplied by $\sec(el)$ to yield the azimuth coordinate offsets that are applied to the azimuth axis encoder. In addition to the fixed tilt given by these coefficients, the mount employs a two-axis tilt meter. The signals from this tilt meter are passed through a 1 Hz low-pass filter and then used as additional corrections, which are applied to the encoders and recorded to disk for use in pointing reconstruction for data analysis. The tilt-meter pointing corrections are a combination of the residual tilt left over from leveling the mount, temporal tilts, and any zero offset of the meter itself. The most significant temporal tilt is an approximately 10 millidegree tilt away from the direction of the Sun during the day caused by the expansion of the sunlit side of the pedestal of the mount.
\begin{table}[h!]
  \begin{center}
    \caption{Boresight Terms.\label{tab:boresight}}
    \begin{tabular}{cc} %
    \toprule 
    Az Terms          & Physical Meaning \\
    \midrule
    $1$               & Collimation Error \\
    $\sin(el)$         & Elevation Axis Orthogonality Error\\
    $\cos(el)$         & Encoder Offset \\
    $\cos(el)\sin(az)$  & Encoder Eccentricity \\
    $\cos(el)\cos(az)$  & Encoder Eccentricity \\
    $\sin(el)\sin(az)$  & Tilt \\
    $\sin(el)\cos(az)$  & Tilt \\
    $\cos(el)\sin(2az)$ & Encoder Eccentricity \tablenotemark{a} \\
    $\cos(el)\cos(2az)$ & Encoder Eccentricity \tablenotemark{a} \\
    $\sin(el)\sin(2az)$ & Azimuth Axis Warp \tablenotemark{a} \\
    $\sin(el)\cos(2az)$ & Azimuth Axis Warp \tablenotemark{a} \\
    \midrule
    \midrule
    El Terms         \\
    \midrule
    $1$               & Collimation Error, Encoder Offset \\
    $\cot(el)$         & Refraction \tablenotemark{b} \\
    $\cos(el)$         & Gravity, Encoder Eccentricity \\
    $\sin(el)$         & Gravity, Encoder Eccentricity \\
    $\cos(2el)$        & Encoder Eccentricity \tablenotemark{a} \\
    $\sin(2el)$        & Encoder Eccentricity \tablenotemark{a} \\
    $\cos(az)$         & Tilt \\
    $\sin(az)$         & Tilt \\
    $\cos(2az)$        & Azimuth Axis Warp \tablenotemark{a} \\
    $\sin(2az)$        & Azimuth Axis Warp \tablenotemark{a} \\
    \bottomrule 
    \end{tabular}
   \tablenotetext{a}{These terms, while included in the model, were found to \\
    be insignificant and have their coefficients set to zero.}
   \tablenotetext{b}{This term, while included in the model, was found to be \\
    degenerate with the gravity terms over the elevation range \\ 
    of the observations and has its coefficient set to zero.}
  \end{center}
\end{table}

\subsection{Boresight Angle Pointing Model}
Since rotating the boresight platform by a given angle as read by the boresight axis encoder does not correspond to the true angle of rotation on the sky, we need a model to correct the boresight angle pointing. The model we use consists of parabolas in commanded boresight angle ($bo$) whose coefficients are themselves parabolas in elevation. This is best visualized as a $3\times 3$ matrix as shown in Table \ref{tab:boresight-angle}.
\begin{table}[h!]
  \begin{center}
    \caption{Boresight Angle Terms.\label{tab:boresight-angle}}
    \begin{tabular} {CCC}     
    \toprule 
    1 & bo & bo^2 \\
    (el-45^{\circ}) & (el-45^{\circ})bo & (el-45^{\circ})bo^2 \\
    (el-45^{\circ})^2 & (el-45^{\circ})^2bo & (el-45^{\circ})^2bo^2 \\
    \bottomrule 
    \end{tabular}
  \end{center}
\end{table}

\subsection{Array Pointing Model}
The array pointing model is described in the instrument frame in what we call the \textit{receiver coordinate system}. This is a spherical coordinate system centered on the equator of the sphere at zero longitude. The principle axes are $X$ and $Y$, where $X$ corresponds to azimuth and $Y$ corresponds to elevation at a boresight angle of zero. Each detector has an offset $\Delta x$ and $\Delta y$ with respect to the origin at array center. Since the fields of view of our telescopes are large, these offsets are computed using spherical trigonometry. The array center itself is not aligned with the boresight of the telescope mount, so we need a pointing model to describe the offset of each telescope's array center with respect to the boresight. While the array center offsets are small, we use spherical trigonometry to describe them for consistency with the detector offsets from the array center. The model uses two coefficients in each of $X$ and $Y$ as shown in Table \ref{tab:array-center}. The elevation dependency is due to gravitational deflections in the optics that change with elevation.
\begin{table}
  \begin{center}
    \caption{Array Center Terms.\label{tab:array-center}}
    \begin{tabular}{cc} %
    \toprule 
    X Terms       & Physical Meaning \\
    \midrule
    $1$           & Collimation Error \\
    $\sin(el-45^{\circ})$  & Gravity \\
    \midrule
    \midrule
    Y Terms       \\
    \midrule
    $1$           & Collimation Error \\
    $\sin(el-45^{\circ})$  & Gravity \\
    \bottomrule 
    \end{tabular}
  \end{center}
\end{table}

\subsection{Pointing Model Data Reduction}
Pointing model data reduction begins with the analysis of a set of drift scans of the Moon during which the telescopes are scanned back and forth at a constant elevation while the Moon rises or sets through their fields of view. This analysis yields the position of each detector on the sky at the point when the Moon is at the beam center along with the mount position at that time and an average boresight angle offset from the commanded boresight angle determined through a minimization of the detector positions with respect to their positions as given by the current array pointing model for each telescope projected onto the sky. These data are then used in an iterative nonlinear least-squares fitting procedure to determine the pointing model coefficients for each telescope. The residual RMS in azimuth, elevation, and boresight angle is typically less than 1 arcminute for a given set of Moon scans. The standard errors of the individual detector offsets from array center are typically less than 10 arcseconds.

\subsection{Pointing Model Usage}\label{sec:pointing_model_usage}
Here we describe the way that the pointing model is used on the telescope mount to solve for the encoder positions and how it is used during data analysis to recover the position of the receiver's array center position from the recorded encoder positions.

\begin{figure}[h]
\centering
\includegraphics[width=0.85\linewidth]{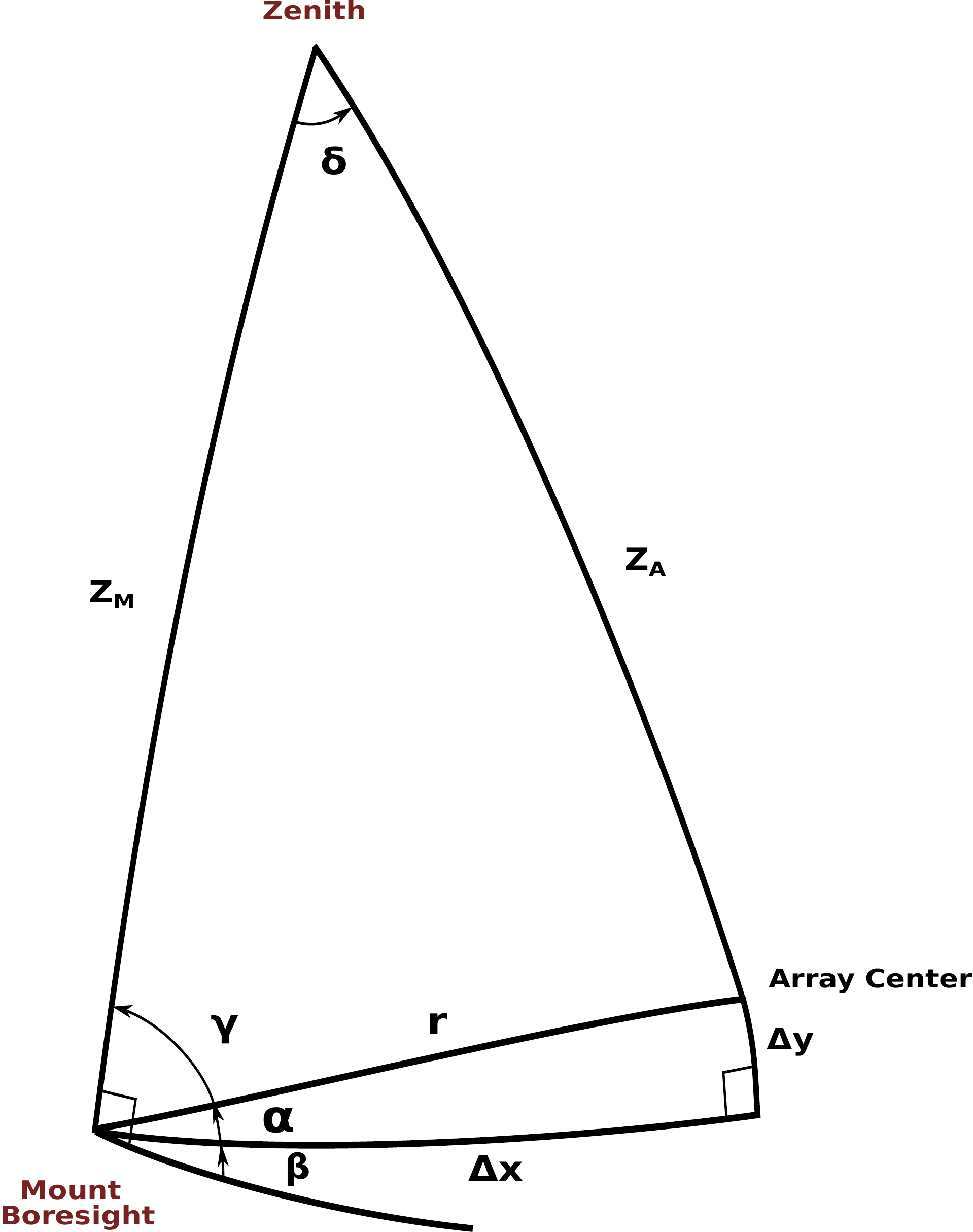}
\caption{Geometry of the offset of array center from the mount boresight showing the symbols used in the equations in Appendix \ref{sec:pointing_model_usage}. Here, $\beta$ is the boresight angle, $\mathrm{r}$ is the spherical distance from the mount boresight to array center, $\Delta x$ and $\Delta y$ are the coordinates of array center with respect to the mount boresight in the receiver coordinate system, and $Z_\mathrm{M}$ and $Z_\mathrm{A}$ are the zenith angles of the mount boresight and array center, respectively.}
\label{fig:array-center-offset}
\end{figure}

\subsubsection{Mount Usage}\label{sec:mount_usage}
On the telescope mount, we are given the desired array center position and boresight angle. First, we calculate the offset of the array center from the boresight in the receiver coordinate system by summing the product of the terms shown in Table \ref{tab:array-center} and their coefficients to derive both the $X$ and $Y$ offsets:
\begin{align}\label{eq:array-center-offsets}
\Delta x &= \sum_{i=1}^{2} a_i f_i(el_\mathrm{M}) \\
\Delta y &= \sum_{i=1}^{2} b_i g_i(el_\mathrm{M}),
\end{align}
where $f_i$, $g_i$ are the $X$ and $Y$ terms in Table \ref{tab:array-center}; $a_i$, $b_i$ are their coefficients; and $el_\mathrm{M}$ is the elevation of the mount boresight, initially set to the desired elevation of array center $el_\mathrm{A}$. Since these offsets are in the receiver coordinate system, they must be projected onto to sky using spherical trigonometry. In this discussion, the symbols shown in Figure \ref{fig:array-center-offset} are used for the relevant quantities. From the properties of the right spherical triangles, we have:
\begin{align}\label{eq:sph-trig1}
\mathrm{r} &= \arccos(\cos(\Delta x)\cos(\Delta y)) \\
\sin(\alpha) &= \sin(\Delta y) / \sin(\mathrm{r}) \\
\cos(\alpha) &= \sin(\Delta x)\cos(\Delta y) / \sin(\mathrm{r}).
\end{align}
From this, we derive $\alpha$ using $\mathrm{atan2}$ for proper quadrant placement. This yields
\begin{align}\label{eq:sph-trig2}
\gamma &= 90^{\circ} - \alpha - \beta,
\end{align}
where $\beta$ is the desired boresight angle. Now we can derive the offsets to the boresight from the law of sines and Napier's analogies
\begin{align}\label{eq:sph-trig3}
\delta &= \arcsin(\sin(\mathrm{r})\sin(\gamma) / \sin(Z_\mathrm{A}) \\
Z_\mathrm{M}  &= 2\arctan\left(\frac{\tan((Z_\mathrm{A} + \mathrm{r})/2)\cos((\gamma + \delta)/2)}{\cos((\gamma - \delta)/2)}\right), \mspace{4mu}\cos(\gamma) \ge 0 \\
Z_\mathrm{M}  &= 2\arctan\left(\frac{\tan((Z_\mathrm{A} - \mathrm{r})/2)\cos((\gamma + \delta))/2}{\cos((\gamma - \delta)/2)}\right), \mspace{4mu}\cos(\gamma) < 0 \\
az_\mathrm{M} &= az_\mathrm{A} - \delta \\
el_\mathrm{M} &= 90^{\circ} - Z_\mathrm{M}.
\end{align}
Since we initially used $el_\mathrm{A}$ to calculate the array center offset, we need to iterate on this. One iteration is sufficient to an accuracy of six decimal places.

Once we have the mount boresight coordinates, we can proceed to calculate the encoder offsets
\begin{align}
\Delta\mspace{2mu}az &= \sec(el_\mathrm{M}) \sum_{i=1}^{11} a_i f_i(az_\mathrm{M}, el_\mathrm{M}) + \tan(el_\mathrm{M})\mathrm{Tilt}_{az} \label{eq:encoder-az-offset} \\
\Delta\mspace{2mu}el &= \sum_{i=1}^{10} b_i g_i(az_\mathrm{M}, el_\mathrm{M}) + \mathrm{Tilt}_{el} \label{eq:encoder-el-offset} \\
\Delta\mspace{2mu}bo &= \sum_{j=1}^{3} \sum_{i=1}^{3} c_{ij} h_{ij}(el_\mathrm{M}, bo) \label{eq:encoder-bo-offset},
\end{align}
where $f_i$, $g_i$ are the $\mathrm{Az}$ and $\mathrm{El}$ terms in Table \ref{tab:boresight}; $h_{ij}$ are the terms in Table \ref{tab:boresight-angle}; and $a_i$, $b_i$, $c_{ij}$ are their coefficients. $\mathrm{Tilt}_{az}$, $\mathrm{Tilt}_{el}$ are the tilt meter corrections. Then,
\begin{align} \label{eq:encoder-positions}
\mathrm{encoder}_{az} &= az_\mathrm{M} + \Delta\mspace{2mu}az \\ 
\mathrm{encoder}_{el} &= el_\mathrm{M} + \Delta\mspace{2mu}el \\
\mathrm{encoder}_{bo} &= bo  + \Delta\mspace{2mu}bo. 
\end{align}

\subsubsection{Pointing Reconstruction Usage}
\begin{figure}[h]
\centering
\includegraphics[width=1\linewidth]{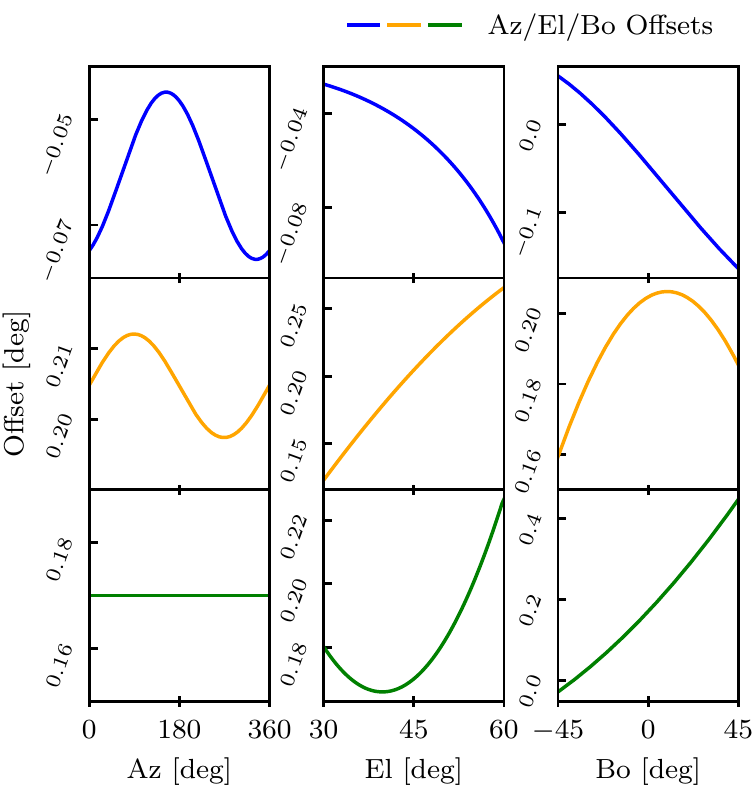}

\caption{Typical pointing corrections for the 40 GHz array. Corrections in azimuth, elevation, and boresight angle are shown in blue, orange, and green, respectively. \textit{Left}: corrections as a function of azimuth at an elevation of $45^{\circ}$ and a boresight angle of $0^{\circ}$. \textit{Middle}: corrections as a function of elevation at an azimuth of $180^{\circ}$ and a boresight angle of $0^{\circ}$. \textit{Right}: corrections as a function of boresight angle at an azimuth of $180^{\circ}$ and an elevation of $45^{\circ}$.}
\label{fig:pointing-corrections}
\end{figure}

Whenever a new pointing model is constructed, a file is stored on disk that contains the new model coefficients. This file is identified by the name of the receiver and the starting date of the model. During data read-in to create spans, the model appropriate for each data package is read from the disk along with the encoder values. The mount boresight coordinates are initially set to the encoder values; then, Equations \ref{eq:encoder-az-offset}, \ref{eq:encoder-el-offset}, and \ref{eq:encoder-bo-offset} are used to calculate the $az$, $el$, and $bo$ encoder offsets. These offsets are subtracted from the encoder values to give an estimate of the mount boresight coordinates $az_\mathrm{M}$, $el_\mathrm{M}$ and the boresight angle $bo$. An estimate of the array center coordinates $az_\mathrm{A}$ and $el_\mathrm{A}$ is derived by following the same procedure as described in Appendix \ref{sec:mount_usage} up through Equation \ref{eq:sph-trig2}. Then the law of cosines is used to derive $Z_\mathrm{A}$
\begin{align}\label{eq:sph-trig4}
Z_\mathrm{A}  &= \arccos(\cos(\mathrm{r})\cos(Z_\mathrm{M}) + \sin(\mathrm{r})\sin(Z_\mathrm{M})\cos(\gamma)).
\end{align}
After calculating $\delta$, the array center coordinates are given by
\begin{align}\label{eq:sph-trig5}
az_\mathrm{A} &= az_\mathrm{M} + \delta \\
el_\mathrm{A} &= 90^{\circ} - Z_\mathrm{A}.
\end{align}
This process is iterated using the new estimate of the mount boresight coordinates. After iterating once, the array center coordinates are converged to six decimal places. Typical pointing corrections for the 40 GHz array are shown in Figure \ref{fig:pointing-corrections}.

\end{document}